\documentclass[twocolumn, onecolappendix]{aastex631}

\usepackage[hyphens]{url}

\shorttitle{Nuclear Environments of TDE Hosts}
\shortauthors{Newsome et al.}

\graphicspath{{./}}
\usepackage{amsmath}
\begin{document}

\title{Resolving the Nuclear Environments of Tidal Disruption Event Host Galaxies within 45 pc}

\newcommand{\LCO}{\affiliation{Las Cumbres Observatory, 6740 Cortona Drive, Suite 102, Goleta, CA 93117-5575, USA}}
\newcommand{\UCSB}{\affiliation{Department of Physics, University of California, Santa Barbara, CA 93106-9530, USA}}
\newcommand{\Iair}
{\affiliation{CIFAR Azrieli Global Scholars program, CIFAR, Toronto, Canada}}
\newcommand{\KITP}{\affiliation{Kavli Institute for Theoretical Physics, University of California, Santa Barbara, CA 93106-4030, USA}}
\newcommand{\UCD}{\affiliation{Department of Physics and Astronomy, University of California, Davis, 1 Shields Avenue, Davis, CA 95616-5270, USA}}
\newcommand{\WIS}{\affiliation{Department of Particle Physics and Astrophysics, Weizmann Institute of Science, 76100 Rehovot, Israel}}
\newcommand{\OKC}{\affiliation{Oskar Klein Centre, Department of Astronomy, Stockholm University, Albanova University Centre, SE-106 91 Stockholm, Sweden}}
\newcommand{\OAPD}{\affiliation{INAF -- Osservatorio Astronomico di Padova, Vicolo dell'Osservatorio 5, I-35122 Padova, Italy}}
\newcommand{\Caltech}{\affiliation{Cahill Center for Astronomy and Astrophysics, California Institute of Technology, Mail Code 249-17, Pasadena, CA 91125, USA}}
\newcommand{\GSFC}{\affiliation{Astrophysics Science Division, NASA Goddard Space Flight Center, Mail Code 661, Greenbelt, MD 20771, USA}}
\newcommand{\UMD}{\affiliation{Joint Space-Science Institute, University of Maryland, College Park, MD 20742, USA}}
\newcommand{\UCB}{\affiliation{Department of Astronomy, University of California, Berkeley, CA 94720-3411, USA}}
\newcommand{\TTU}{\affiliation{Department of Physics, Texas Tech University, Box 41051, Lubbock, TX 79409-1051, USA}}
\newcommand{\STScI}{\affiliation{Space Telescope Science Institute, 3700 San Martin Drive, Baltimore, MD 21218-2410, USA}}
\newcommand{\UT}{\affiliation{University of Texas at Austin, 1 University Station C1400, Austin, TX 78712-0259, USA}}
\newcommand{\IoA}{\affiliation{Institute of Astronomy, University of Cambridge, Madingley Road, Cambridge CB3 0HA, UK}}
\newcommand{\QUB}{\affiliation{Astrophysics Research Centre, School of Mathematics and Physics, Queen's University Belfast, Belfast BT7 1NN, UK}}
\newcommand{\IPAC}{\affiliation{Spitzer Science Center, California Institute of Technology, Pasadena, CA 91125, USA}}
\newcommand{\JPL}{\affiliation{Jet Propulsion Laboratory, California Institute of Technology, 4800 Oak Grove Dr, Pasadena, CA 91109, USA}}
\newcommand{\Southampton}{\affiliation{Department of Physics and Astronomy, University of Southampton, Southampton SO17 1BJ, UK}}
\newcommand{\LANL}{\affiliation{Space and Remote Sensing, MS B244, Los Alamos National Laboratory, Los Alamos, NM 87545, USA}}
\newcommand{\Tsinghua}{\affiliation{Physics Department and Tsinghua Center for Astrophysics, Tsinghua University, Beijing, 100084, People's Republic of China}}
\newcommand{\NAOC}{\affiliation{National Astronomical Observatory of China, Chinese Academy of Sciences, Beijing, 100012, People's Republic of China}}
\newcommand{\Itagaki}{\affiliation{Itagaki Astronomical Observatory, Yamagata 990-2492, Japan}}
\newcommand{\Einstein}{\altaffiliation{Einstein Fellow}}
\newcommand{\Hubble}{\altaffiliation{Hubble Fellow}}
\newcommand{\CfA}{\affiliation{Center for Astrophysics \textbar{} Harvard \& Smithsonian, 60 Garden Street, Cambridge, MA 02138-1516, USA}}
\newcommand{\UA}{\affiliation{Steward Observatory, University of Arizona, 933 North Cherry Avenue, Tucson, AZ 85721-0065, USA}}
\newcommand{\MPIA}{\affiliation{Max-Planck-Institut f\"ur Astrophysik, Karl-Schwarzschild-Stra\ss{}e 1, D-85748 Garching, Germany}}
\newcommand{\DSFP}{\altaffiliation{LSSTC Data Science Fellow}}
\newcommand{\HCO}{\affiliation{Harvard College Observatory, 60 Garden Street, Cambridge, MA 02138-1516, USA}}
\newcommand{\Carnegie}{\affiliation{Observatories of the Carnegie Institute for Science, 813 Santa Barbara Street, Pasadena, CA 91101-1232, USA}}
\newcommand{\TAU}{\affiliation{School of Physics and Astronomy, Tel Aviv University, Tel Aviv 69978, Israel}}
\newcommand{\Edinburgh}{\affiliation{Institute for Astronomy, University of Edinburgh, Royal Observatory, Blackford Hill EH9 3HJ, UK}}
\newcommand{\Birmingham}{\affiliation{Birmingham Institute for Gravitational Wave Astronomy and School of Physics and Astronomy, University of Birmingham, Birmingham B15 2TT, UK}}
\newcommand{\Bath}{\affiliation{Department of Physics, University of Bath, Claverton Down, Bath BA2 7AY, UK}}
\newcommand{\CTIO}{\affiliation{Cerro Tololo Inter-American Observatory, National Optical Astronomy Observatory, Casilla 603, La Serena, Chile}}
\newcommand{\Potsdam}{\affiliation{Institut f\"ur Physik und Astronomie, Universit\"at Potsdam, Haus 28, Karl-Liebknecht-Str. 24/25, D-14476 Potsdam-Golm, Germany}}
\newcommand{\INPE}{\affiliation{Instituto Nacional de Pesquisas Espaciais, Avenida dos Astronautas 1758, 12227-010, S\~ao Jos\'e dos Campos -- SP, Brazil}}
\newcommand{\UNC}{\affiliation{Department of Physics and Astronomy, University of North Carolina, 120 East Cameron Avenue, Chapel Hill, NC 27599, USA}}
\newcommand{\Ohio}{\affiliation{Astrophysical Institute, Department of Physics and Astronomy, 251B Clippinger Lab, Ohio University, Athens, OH 45701-2942, USA}}
\newcommand{\AAS}{\affiliation{American Astronomical Society, 1667 K~Street NW, Suite 800, Washington, DC 20006-1681, USA}}
\newcommand{\MMT}{\affiliation{MMT and Steward Observatories, University of Arizona, 933 North Cherry Avenue, Tucson, AZ 85721-0065, USA}}
\newcommand{\Geneva}{\affiliation{ISDC, Department of Astronomy, University of Geneva, Chemin d'\'Ecogia, 16 CH-1290 Versoix, Switzerland}}
\newcommand{\IUCAA}{\affiliation{Inter-University Center for Astronomy and Astrophysics, Post Bag 4, Ganeshkhind, Pune, Maharashtra 411007, India}}
\newcommand{\CMU}{\affiliation{Department of Physics, Carnegie Mellon University, 5000 Forbes Avenue, Pittsburgh, PA 15213-3815, USA}}
\newcommand{\NAOJ}{\affiliation{Division of Science, National Astronomical Observatory of Japan, 2-21-1 Osawa, Mitaka, Tokyo 181-8588, Japan}}
\newcommand{\IfA}{\affiliation{Institute for Astronomy, University of Hawai`i, 2680 Woodlawn Drive, Honolulu, HI 96822-1839, USA}}
\newcommand{\UCSC}{\affiliation{Department of Astronomy and Astrophysics, University of California, Santa Cruz, CA 95064-1077, USA}}
\newcommand{\Purdue}{\affiliation{Department of Physics and Astronomy, Purdue University, 525 Northwestern Avenue, West Lafayette, IN 47907-2036, USA}}
\newcommand{\Princeton}{\affiliation{Department of Astrophysical Sciences, Princeton University, 4 Ivy Lane, Princeton, NJ 08540-7219, USA}}
\newcommand{\Moore}{\affiliation{Gordon and Betty Moore Foundation, 1661 Page Mill Road, Palo Alto, CA 94304-1209, USA}}
\newcommand{\Durham}{\affiliation{Department of Physics, Durham University, South Road, Durham, DH1 3LE, UK}}
\newcommand{\JHU}{\affiliation{Department of Physics and Astronomy, The Johns Hopkins University, 3400 North Charles Street, Baltimore, MD 21218, USA}}
\newcommand{\Toronto}{\affiliation{David A.\ Dunlap Department of Astronomy and Astrophysics, University of Toronto,\\ 50 St.\ George Street, Toronto, Ontario, M5S 3H4 Canada}}
\newcommand{\Duke}{\affiliation{Department of Physics, Duke University, Campus Box 90305, Durham, NC 27708, USA}}
\newcommand{\NCU}{\affiliation{Graduate Institute of Astronomy, National Central University, 300 Jhongda Road, 32001 Jhongli, Taiwan}}
\newcommand{\Columbia}{\affiliation{Department of Physics and Columbia Astrophysics Laboratory, Columbia University, Pupin Hall, New York, NY 10027, USA}}
\newcommand{\Flatiron}{\affiliation{Center for Computational Astrophysics, Flatiron Institute, 162 5th Avenue, New York, NY 10010-5902, USA}}
\newcommand{\CIERA}{\affiliation{Center for Interdisciplinary Exploration and Research in Astrophysics and Department of Physics and Astronomy, \\Northwestern University, 1800 Sherman Avenue, 8th Floor, Evanston, IL 60201, USA}}
\newcommand{\GeminiObs}{\affiliation{Gemini Observatory, 670 North A`ohoku Place, Hilo, HI 96720-2700, USA}}
\newcommand{\Keck}{\affiliation{W.~M.~Keck Observatory, 65-1120 M\=amalahoa Highway, Kamuela, HI 96743-8431, USA}}
\newcommand{\UW}{\affiliation{Department of Astronomy, University of Washington, 3910 15th Avenue NE, Seattle, WA 98195-0002, USA}}
\newcommand{\DiRAC}{\altaffiliation{DiRAC Fellow}}
\newcommand{\USask}{\affiliation{Department of Physics \& Engineering Physics, University of Saskatchewan, 116 Science Place, Saskatoon, SK S7N 5E2, Canada}}
\newcommand{\Thacher}{\affiliation{Thacher School, 5025 Thacher Road, Ojai, CA 93023-8304, USA}}
\newcommand{\Rutgers}{\affiliation{Department of Physics and Astronomy, Rutgers, the State University of New Jersey,\\136 Frelinghuysen Road, Piscataway, NJ 08854-8019, USA}}
\newcommand{\FSU}{\affiliation{Department of Physics, Florida State University, 77 Chieftan Way, Tallahassee, FL 32306-4350, USA}}
\newcommand{\Melbourne}{\affiliation{School of Physics, The University of Melbourne, Parkville, VIC 3010, Australia}}
\newcommand{\ASTROthreeD}{\affiliation{ARC Centre of Excellence for All Sky Astrophysics in 3 Dimensions (ASTRO 3D)}}
\newcommand{\Stromlo}{\affiliation{Mt.\ Stromlo Observatory, The Research School of Astronomy and Astrophysics, Australian National University, ACT 2601, Australia}}
\newcommand{\NCPAS}{\affiliation{National Centre for the Public Awareness of Science, Australian National University, Canberra, ACT 2611, Australia}}
\newcommand{\TAMU}{\affiliation{Department of Physics and Astronomy, Texas A\&M University, 4242 TAMU, College Station, TX 77843, USA}}
\newcommand{\Mitchell}{\affiliation{George P.\ and Cynthia Woods Mitchell Institute for Fundamental Physics \& Astronomy, College Station, TX 77843, USA}}
\newcommand{\ESO}{\affiliation{European Southern Observatory, Alonso de C\'ordova 3107, Casilla 19, Santiago, Chile}}
\newcommand{\ICE}{\affiliation{Institute of Space Sciences (ICE, CSIC), Campus UAB, Carrer
de Can Magrans, s/n, E-08193 Barcelona, Spain}}
\newcommand{\IEEC}{\affiliation{Institut d'Estudis Espacials de Catalunya, Gran Capit\`a, 2-4, Edifici Nexus, Desp.\ 201, E-08034 Barcelona, Spain}}
\newcommand{\Warwick}{\affiliation{Department of Physics, University of Warwick, Gibbet Hill Road, Coventry CV4 7AL, UK}}
\newcommand{\Macquarie}{\affiliation{School of Mathematical and Physical Sciences, Macquarie University, NSW 2109, Australia}}
\newcommand{\AAARC}{\affiliation{Astronomy, Astrophysics and Astrophotonics Research Centre, Macquarie University, Sydney, NSW 2109, Australia}}
\newcommand{\Capodimonte}{\affiliation{INAF -- Capodimonte Astronomical Observatory, Salita Moiariello 16, I-80131 Napoli, Italy}}
\newcommand{\INFNNapoli}{\affiliation{INFN -- Napoli, Strada Comunale Cinthia, I-80126 Napoli, Italy}}
\newcommand{\ICRANet}{\affiliation{ICRANet, Piazza della Repubblica 10, I-65122 Pescara, Italy}}
\newcommand{\UVA}{\affiliation{Department of Astronomy, University of Virginia, Charlottesville, VA 22904, USA}}
\newcommand{\UIUC}{\affiliation{Department of Astronomy, University of Illinois, 1002 W. Green Street, Urbana, IL 61801, USA}}
\newcommand{\Leiden}{\affiliation{Leiden Observatory, Leiden University, Postbus 9513, 2300 RA, Leiden, The Netherlands}}
\newcommand{\HUJ}{\affiliation{Racah Institute of Physics, The Hebrew University, 91904 Jerusalem, Israel}}
\newcommand{\UWM}{\affiliation{University of Wisconsin-Madison, Department of Astronomy, 475 N Charter St, Madison, WI 53703, USA}}
\newcommand{\Polar}{\affiliation{Polar Research Institute of China, 451 Jinqiao Road, Pudong, Shanghai 200136, China}}
\newcommand{\USTC}{\affiliation{Department of Astronomy, University of Science and Technology of China, Hefei, Anhui 230026, China}}

\author[0000-0001-9570-0584]{Megan Newsome}
\UT
\author[0000-0001-7090-4898]{Iair Arcavi}
\TAU
\author[0000-0002-4235-7337]{K. Decker French}
\UIUC
\author[0000-0001-5807-7893]{Curtis McCully}
\LCO
\author[0000-0001-6047-8469]{Ann Zabludoff}
\UA
\author[0000-0002-4337-9458]{Nicholas Stone}
\HUJ
\UWM
\author[0000-0002-3859-8074]{Sjoert van Velzen}
\Leiden
\author{Tinggui Wang}
\USTC

\begin{abstract}

Using HST/STIS observations, we present the highest-spatial-resolution spectroscopic study to date of four tidal disruption event (TDE) host galaxies, with the best observed being the post-starburst (PSB) host of ASASSN-14li. The stellar population of ASASSN-14li's host, within 44 pc of the nucleus, reveals a younger recent starburst ($\sim340 \, \mathrm{Myr}$) compared to the population at an offset radius of 88 pc that excludes the nucleus ($\sim550 \, \mathrm{Myr}$), a radial age gradient suggesting gas inflows from a minor merger. We estimate a stellar density of $\sim5900 \pm 800 \, M_\odot / \mathrm{pc}^3$ within 30 pc of the nucleus of ASASSN-14li's host, exceeding densities expected for nuclear star clusters. High-ionization ``coronal" emission lines, $[$\ion{Fe}{6}$]$ $\lambda 5677$, $[$\ion{Fe}{7}$]$ $\lambda 6087$, and $[$\ion{Fe}{10}$]$ $\lambda 6375$, are also detected within the nuclear spectra of the hosts of ASASSN-14li and PTF09ge, importantly alongside the non-detection of [\ion{O}{3}] at the same scale. We similarly do not detect [\ion{O}{3}] in the nuclear region of ASASSN-14ae's host despite its presence in the SDSS spectrum. The different ionization radiation levels detected at various radii from TDE host nuclei may indicate echoes of earlier accretion episodes, including, potentially, a prior TDE. We posit that a minor merger driving gas inflow to the nucleus could drive the enhanced TDE rates in post-starburst galaxies, inducing variation in nuclear gas properties and star formation history on $<$150 pc scales in TDE hosts.
\end{abstract}


\section{Introduction} \label{sec:intro}

\begin{deluxetable*}{cccccc}[t!]\label{table:observations}
    \caption{HST/STIS Observations of TDE Hosts from GO-14717}
    \tablehead{
    \colhead{TDE} & \colhead{Grating} & \colhead{Exp. Time (s)} & 
    \colhead{HST Filing Name} & \colhead{Size of 0.2$''$ (pc)} & \colhead{Min. Extraction Radius (pc)} }
    \startdata
        ASASSN-14li & G430L & 675 & odah15010 & 88 & 44\\
        ASASSN-14li & G750L & 675 & odah15020 & 88 & 44\\
        ASASSN-14ae & G430L & 4470 & odah14010 & 178 & 66\\
        ASASSN-14ae & G750L & 2840 & odah14020 & 178 & 66\\
        PTF09ge & G430L & 1815 & odah13010 & 257 & 102\\
        PTF09ge & G750L & 2950 & odah13020 & 257 & 102\\
        iPTF15af & G430L & 1685 & odah16010 & 310 & 124\\
        iPTF15af & G750L & 2815 & odah16020 & 310 & 124\\
    \enddata
\end{deluxetable*}

When stars approach too close to a supermassive black hole (SMBH) in a galaxy center, the tidal forces overcome the star's self-gravity and create streams of stellar debris on a path toward the SMBH \citep{1975Natur.254..295H, Rees1988}. The stream of material that accretes onto the SMBH can subsequently produce an X-ray, ultraviolet, and/or optical flare over the timescale of months to years, and many cases also show late-time radio flares \citep[see e.g.][]{Alexander2020, Cendes2024}. Dozens of these tidal disruption events (TDEs) have now been observed with a wide variety of surprising characteristics \citep[e.g.][]{Gezari2012, Gezari2021, VanVelzen21, Nicholl2022, Hammerstein2023, Clark2023, Hinkle2023, Newsome2024b}.

One puzzling feature of UV/optical TDEs is their preference for post-starburst host galaxies \citep[PSBs;][]{Arcavi2014, French2016, French2020}. These galaxies are very rare and lack ongoing star formation, while still showing stellar populations indicative of star formation that ceased abruptly up to $\sim$1.4 Gyr ago \citep{DresslerGunn1983}. Optical TDE rates are enhanced by $\sim$10-30$\times$ in quiescent, Balmer-strong and post-starburst galaxies \citep{French2016, LawSmith2017, Graur2018, Hammerstein2021}. The magnitude of this rate enhancement has declined from initial estimates as the sample of TDEs has grown to include a broader range of host types, especially when considering non-optical TDEs, indicating that the perceived effect of post-starburst galaxies on becoming TDE hosts may be influenced by selection effects \citep[e.g.][]{Hammerstein2021, Sazonov2021, Masterson2024}. Still, the over-representation of the otherwise-rare post-starburst and quiescent Balmer strong (QBS) galaxies among optically-identified TDEs persists. How the small-scale dynamics of a galaxy's nucleus are connected to the large-scale dynamics of the galaxy’s total star formation history remains to be discerned.

At the same time, \cite{French2017} and \cite{French2020} found that the host of ASASSN-14li is entirely bulge-like, fit best by a S\'ersic profile with index $\approx$3 combined with a blue (F438W - F625W = 0.09) point source within 30pc of the SMBH. All four hosts studied photometrically in \cite{French2020}, the same studied spectroscopically in this work, were found to have higher central surface brightnesses than comparable early-type galaxies and lacked obvious strong merger signatures such as asymmetries.

To discern the cause of the enhanced TDE rates in post-starburst galaxies, higher-resolution spectroscopic analysis is needed to measure differences in stellar populations at small scales ($<$100 pc from the nucleus) and detect or localize the source of emission lines seen from ground-based spectra. Here we present a resolved spatial characterization of the stellar population as a function of distance from the galactic nucleus for the PSB host of ASASSN-14li using \textit{Hubble Space Telescope} (HST) spectroscopy. We also provide the nuclear spectra of three other TDE hosts observed with HST/STIS and discuss their attributes. We review our methods of data extraction and reduction in \S\ref{sec:methods}, review the results of the data in \S \ref{sec:results}, discuss the implications of the properties of the circumnuclear environments in \S\ref{sec:discussion}, and summarize  in \S\ref{sec:conclusion}. Throughout this work we adopt the \cite{2020A&A...641A...6P} cosmology with H$_0$ = 67.7 km s$^{-1}$ Mpc$^{-1}$.

\section{Observations and Methods} \label{sec:methods}

\subsection{The Sample}
The four TDE hosts selected for HST spectroscopy were the nearest targets studied in \cite{French2016}: ASASSN-14li \citep[z=0.0206;][]{Holoien2016}, ASASSN-14ae \citep[z=0.0436;][]{Holoien2014}, PTF09ge \citep[0.064;][]{Arcavi2014}, and iPTF15af \citep[0.079;][]{Blagorodnova2019}. The host of ASASSN-14li is a post-starburst galaxy, while those of ASASSN-14ae and iPTF15af are quiescent Balmer strong (QBS) with prior weak starbursts, and that of PTF09ge is a quiescent early type according to their archival spectra \citep{French2016}. Approximately 40\% of known UV-optical TDEs have QBS hosts, including post-starburst galaxies, corresponding to an over-representation factor of 16x given the rarity of QBS galaxies \citep{French2020b, Hammerstein2021}. Our sample was chosen for being among the closest TDE hosts at the time of selection, but with 3/4 being QBS or post-starburst hosts, it is also a useful sample to investigate the causes of this over-representation.

ASASSN-14li, one of the most well-studied TDEs to date \citep{2015Natur.526..542M, Holoien2016, 2016Sci...351...62V, Alexander2016, JiangN2016, Kara2018, 2018ApJ...856....1P}, occurred in a host found by \cite{French2016, French2017, French2020} to be dominated by a bulge (with $M_{\text{bulge}} = 10^{9.6} M_{\odot}$ while $M_{*} = 10^{9.7} M_{\odot}$). This host was also found to have a higher stellar mass density at the 30pc scale than most early-type galaxies with comparable total stellar mass, while its low asymmetry and lack of tidal features seen in the continuum bands implied that it could not have experienced a recent gas-rich major ($\sim$1:1--1:3 mass ratio) merger, though a minor merger ($\sim$1:3--1:5) could still explain the extended ionized features seen by \cite{Prieto2016}. ASASSN-14li's host has also been studied via integral field spectroscopy at different distances from the nuclear region, through which extended emission line filaments ($\geq 5$kpc from the SMBH) were uncovered \citep{Prieto2016}. These emitting regions are streams of ionized gas whose narrow line widths require a photoionization origin, potentially indicating that the host had AGN activity triggered by a recent merger. It has since been posited that such extended emission line regions may in fact be the result of another TDE in the host's past, as PSB galaxies are predisposed to higher rates of TDE activity, and the long-term ionizing flux from a late TDE plateau phase in the UV can impact the host for $\sim$10,000 years \citep{Mummery2025}.

The next closest TDE host, that of ASASSN-14ae, also has higher central brightness concentration than most early-type galaxies while showing little asymmetry expected of major mergers, and similar features were found in iPTF15af's host, which is a barred lenticular galaxy \citep{French2020}. The host of PTF09ge, also a barred lenticular, is otherwise a quiescent early-type galaxy, but with a blue star-forming ring at a radius of $\sim$3kpc from its nucleus; PTF09ge's host is especially centrally concentrated and was best-fit by a central point source in addition to a S\`ersic profile, similar to ASASSN-14li \citep{French2020}.

\subsection{Observations and Data Processing}

Slit spectroscopy of the four TDE hosts was taken during HST Cycle 24 (PI: I. Arcavi, Proposal ID 14717). The observations were taken using the STIS with gratings G430L and G750L. The slit used has an aperture of 52 arcsec $\times$ 0.2 arcsec. Details of the observations are given in Table \ref{table:observations}. Bias, flat-fielding, and cosmic-ray rejection were automatically performed prior to download from the Mikulski Archive for Space Telescopes\footnote{\url{https://mast.stsci.edu/hlsp/}}. We first applied a correction to the charge transfer inefficiencies in each 2D image using the package \texttt{stis\_cti}\footnote{\url{https://pythonhosted.org/stis_cti/}}. We then corrected the G750L grating's intrinsic fringing via fringe flats obtained with each observation using the module \texttt{defringe} from the package \texttt{STISTOOLS}\footnote{\url{https://stistools.readthedocs.io/}}. The spectra files were then processed using the STIStools \texttt{X1D} script to extract 1D spectra. We eliminated bad pixels using the data-quality flags 16 (high dark rate) and 512 (bad reference pixel). We used a 4-pixel width for each extraction to obtain $\sim80$\% of the flux available from the brightest part of the trace, due to the PSF and LSF of each grating. This is the smallest extraction width, and therefore smallest physical scale, that we can resolve with the 52x0.2$\arcsec$ slit\footnote{\url{https://hst-docs.stsci.edu/stisihb/}}. All the HST data in this paper can be found on MAST: \dataset[10.17909/nd4f-5f64]{http://dx.doi.org/10.17909/nd4f-5f64}.

For our closest host, that of ASASSN-14li, we extracted spectra along the trace of maximum flux corresponding to the host nucleus, as well as at 4 pixels offset in each direction from the nucleus trace, totaling 3 extractions. 
Due to the host's spherical symmetry, we averaged the flux from two extractions taken at $\pm$4-pixel offsets to improve the signal for the offset region. Thus we obtained two total spectra for the host of ASASSN-14li: a nuclear region and an offset region. The redshift of the host and the resolution of STIS combine such that the nucleus extraction includes the flux projected within a 44 pc radius of the central SMBH, while the offset averaged spectrum has a mean projected distance from the SMBH of 88 pc.

For the other three hosts, the redshifts are too high to extract spectroscopy from regions offset from the nucleus as the S/N is too low. We extracted only the traces corresponding to the nucleus with similar 4-pixel extraction widths, corresponding to regions projected within 66 pc for ASASSN-14ae, 102 pc for PTF09ge, and 124 pc for iPTF15af.

\subsection{Star Formation History Analysis}

To infer details of the stellar population and circumnuclear environment on the smallest scales yet discerned for a post-starburst TDE host, we use \texttt{BAGPIPES} \citep{Carnall2018} to find the stellar population ages and masses of both starbursts (old and young) for ASASSN-14li's host. \texttt{BAGPIPES}  fits spectroscopic models of galaxies to observed photometry in order to estimate the host's stellar mass and star formation history (SFH), metallicity, dust content, and age. We follow the setup from \cite{French2017} for fitting the SFHs of ASASSN-14li's old and young starbursts at the nucleus. We use the stellar synthesis models as outlined in \cite{BC03} and updated in 2016\footnote{\url{http://www.bruzual.org/~gbruzual/bc03/Updated_version_2016/}} assuming a \cite{Chabrier2003} initial mass function. The star formation history is modeled as an old stellar population following a delayed exponential and a young population with a simple exponential history. We simultaneously model the correlated flux uncertainty with three uniform-prior terms corresponding to a second-order Chebyshev polynomial. The priors on metallicity and correlated flux calibration parameters are Gaussian, while the priors on the age of the delayed exponential population and the noise of the spectrum are uniform in log-space; all other parameters are fit with flat priors. The metallicity prior is set by the relation from \cite{Gallazzi2005}. We use the \cite{Calzetti2000} dust law with a uniform prior on extinction. We allow the redshift to vary within a narrow range ($\Delta z < 0.005$) of the spectroscopically determined value $z=0.0206$.

\section{Results} \label{sec:results}

\begin{figure*}[t!]
    \centering
    \includegraphics[width=0.82\linewidth]{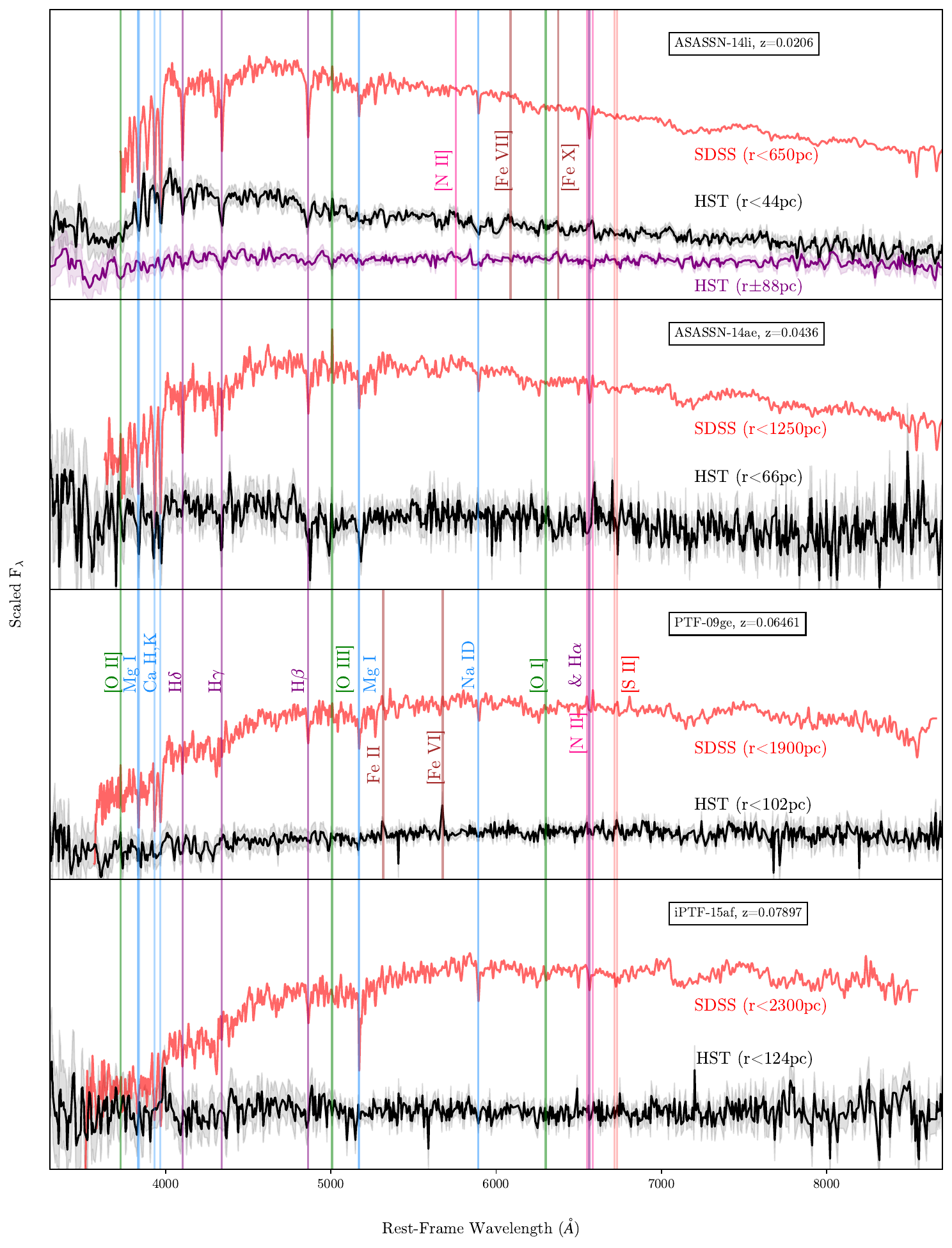}
    \caption{The nuclear spectrum of each TDE host as taken by HST (black, with errors ranges shown in gray) compared to the SDSS spectrum of each host (red). Each panel is a different host: ASASSN-14li (first/top), ASASSN-14ae (second), PTF09ge (third), iPTF15af (fourth/bottom). For ASASSN-14li (top panel) we also show the averaged spectrum from the HST extractions at $\pm$88 pc (purple, with error regions shown in light purple). Lines either expected based on ground-based spectra, or newly identified from these data, are marked with vertical lines and labeled in the third panel. Specific lines identified only for one source, as discussed in \S \ref{sec:results}, are labeled with vertical lines in the corresponding panel (forbidden lines in ASASSN-14li, top panel; iron lines in PTF09ge, third panel). There are several differences between the archival SDSS spectra and the nuclear spectrum of each host from HST, most notably the identification of forbidden lines in both ASASSN-14li and PTF09ge, as well as the lack of nuclear oxygen emission lines in ASASSN-14ae in contrast with its SDSS spectrum.} 
    \label{fig:asassn14li_nucleus_offsets}
\end{figure*}

All four host spectra are shown in Figure \ref{fig:asassn14li_nucleus_offsets}, with the spectrum of ASASSN-14li's host at the nucleus (with flux from within the inner 44 pc), alongside the averaged spectra extracted from the offset regions centered at $\pm$88 pc, in the top panel. We review the results from the star formation history (SFH) fitting of ASASSN-14li's host nuclear and offset spectra, and the consequences of the best-fit results on the density profile of the host. We then detail the detections (and non-detections) of noteworthy emission and absorption lines in the other three TDE hosts observed, particularly in contrast with lines detected in SDSS archival spectra with substantially larger apertures of extraction.

\subsection{A Young Nucleus in ASASSN-14li's Host}
\label{sec:host_fits}

\begin{figure}
    \centering
    \includegraphics[scale=0.33]{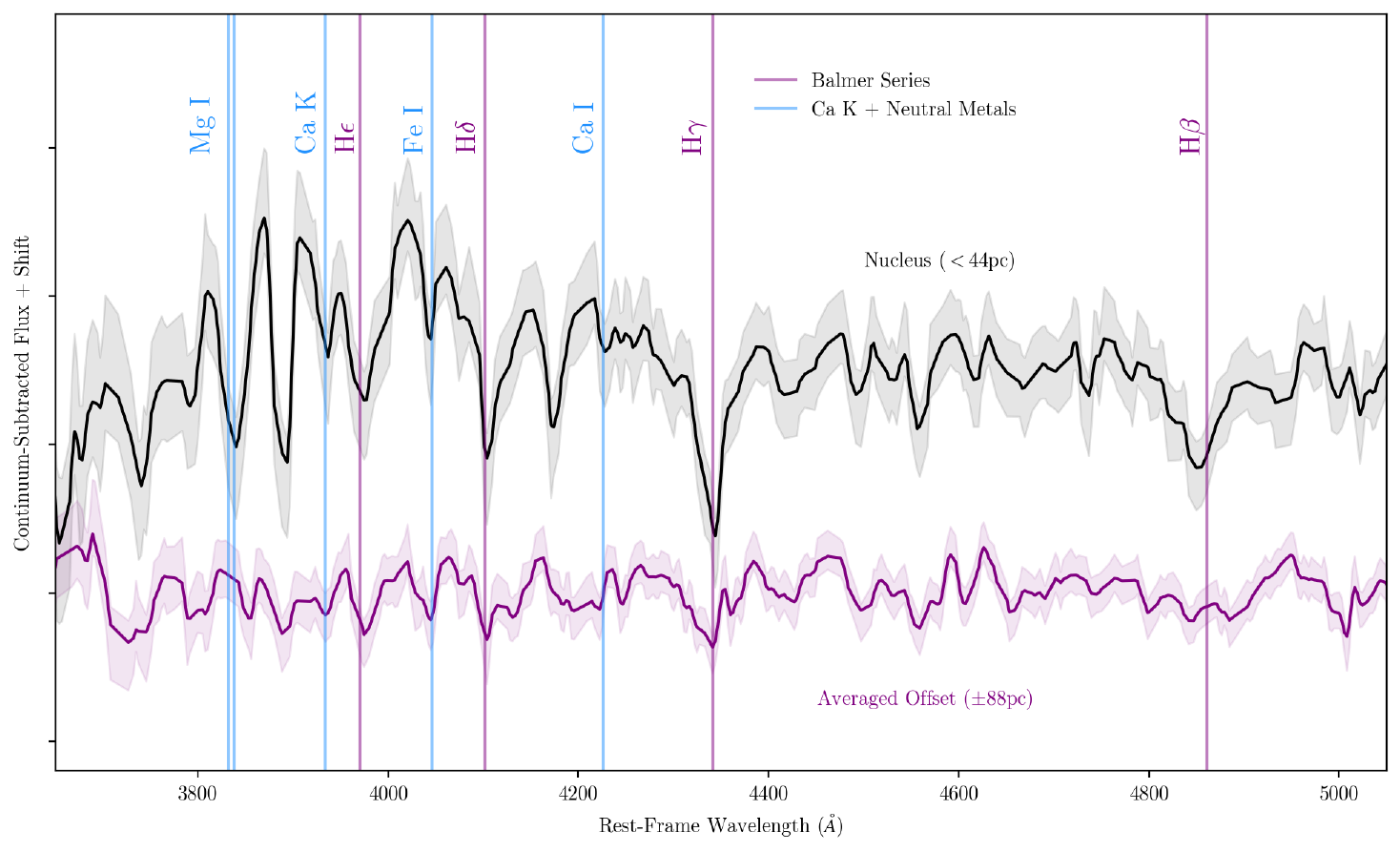}
    \includegraphics[width=0.98\linewidth]{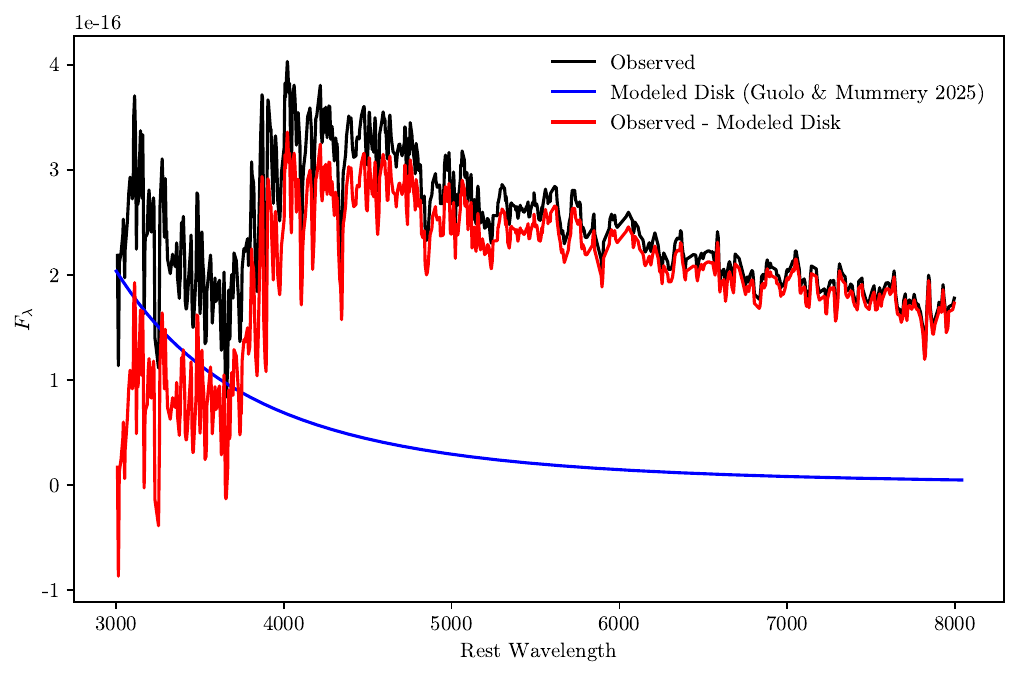}
    \caption{Top: A close-up of the nuclear spectrum of ASASSN-14li (with an extraction radius within 44 pc of the host center, in black with error ranges in gray) and the averaged spectrum from the extractions at $\pm$88 pc (purple, with error ranges in light purple), both smoothed to visually emphasize the similarities in absorption features. We label the key lines used to analyze the stellar ages: Ca K $\lambda$3933 and neutral metals Mg I $\lambda\lambda$3832,3838, Fe I $\lambda$4046, and Ca I $\lambda$4226 (blue vertical lines); and the Balmer series.
    Bottom: The late-time plateau of ASASSN-14li likely arises from an accretion disk (that was either obscured or only recently formed) as modeled by \cite{GuoloMummery2025} and reproduced here in blue. The black line shows our original nuclear spectrum as observed, while the red line shows how the spectrum of the host nucleus changes without the disk contribution. This does not significantly change the overall flux redward of 4000\AA  which defines the strength of the absorption lines used to determine star formation histories with \texttt{BAGPIPES}, however, the strong disk contribution on the bluest end does give the nucleus a steeper upturn blueward of the 4000\AA break, which can still impact stellar population age and mass estimates. We use the resulting subtracted spectrum in red for our \texttt{BAGPIPES} fits. } 
    \label{fig:bagpipes_fit}
\end{figure}

The age and mass of starbursts at different radii from the nucleus of ASASSN-14li's host can help discern whether the TDE rate of PSB hosts is enhanced by A-stars becoming giants or by gas funneling toward the SMBH after a merger. We use \texttt{BAGPIPES} to fit for: the old stellar population's age and mass; the young stellar population's age (time since the peak SFR), mass, metallicity, and timescale of decline $\tau$; dust extinction; ionization parameter $\log$U; velocity dispersion; redshift; and the aforementioned flux calibration and noise parameters. After initially fitting for the calibration and noise parameters, we use their converged values in subsequent fits to limit the parameter space to only 10 values. 

It has been shown that ASASSN-14li has a late-time ``plateau'' primarily in the UV (alongside a steady decline in a delayed X-ray flare) that has been well-modeled as contribution from an accretion disk that has either formed after, or became unobscured after, the rise and fall of the UV-optical lightcurve \citep{2024MNRAS.527.2452M, GuoloMummery2025}. Thus, before fitting, we account for ASASSN-14li's late-time plateau in the rest-frame wavelength range of our nuclear spectrum by reproducing the disk emission estimated by \cite{GuoloMummery2025} using their publicly available \texttt{diskSED}\footnote{\url{https://github.com/muryelgp/diskSED/tree/main/diskSED}} code and their fitting results for the epoch overlapping with our observations (their ``Epoch 3''). Even though the star formation histories are largely determined by the depth of the stellar absorption lines from the continuum between 4000 and 5000 \AA, as opposed to the total flux overall (the calibration offsets that we fit as Chebyshev polynomial coefficients encompass more than 10\% adjustments to the flux), the accretion disk's stronger contribution on the bluest side of our spectrum fundamentally changes the shape of the 4000\AA\, break (see bottom portion of Figure \ref{fig:bagpipes_fit}).

Thus we run the \texttt{BAGPIPES} fitting method with the disk-subtracted spectrum as our nuclear region. The corner plot for fits to both the unsubtracted and disk-subtracted spectra are in the Appendix Figures \ref{fig:bagpipes_corner} and \ref{fig:bagpipes_corner_subtracted}. We find that that the subtracted spectrum is best-fit with a younger starburst population of age 340 Myr, older than that of the unsubtracted spectrum (230 Myr) but still notably younger than the post-starburst age of $\sim$550 Myr found via fits to our offset spectrum centered at $\pm$88 pc as well as fits to the SDSS spectrum \citep{French2017}. We find a total stellar mass within the nuclear region of $\log$(M$_{*}$/M$_{\odot}$)=8.58$^{+0.094}_{-0.085}$. We list all the best-fit parameters from \texttt{BAGPIPES} in Table \ref{table:bagpipes}. These results show that the residual UV light from TDEs at least a $\sim$year after the UV/optical peak, presumably from the delayed-onset or late-uncovering of an accretion disk, can bias star formation history studies of TDE host galaxies to find younger post-starburst populations than may be intrinsic. Contributions from the late-time plateau should thus be accounted for in future analyses of TDE hosts where possible, particularly when relying on blue spectral properties.

\begin{deluxetable*}{ccccc}[t!]\label{table:bagpipes}
    \caption{Fitting ranges and best-fit values of the fit paramters from BAGPIPES of the innermost nuclear region of ASASSN-14li's host. The first (old) starburst of the host is modeled as a delayed exponential power law, while the second (younger) starburst of the host is modeled as an exponential power law. The order of the rows are as follows: Age of the first starburst, mass of the first starburst, age of the second starburst, mass of the second starburst, the values of $\tau$ in the exponential power law, metallicity from the second starburst, ionization parameter, redshift, velocity dispersion and dust extinction.}
    \tablehead{
    \colhead{Parameter} & \colhead{Fitting Range} & \colhead{Best-Fit to Nucleus} & \colhead{Best-Fit to Nucleus} & \colhead{Best-Fit to Offset Region} \\ \colhead{} & \colhead{} & \colhead{(Disk-Subtracted)} & \colhead{(Unsubtracted)} & \colhead{} }
    \startdata
    Age$_\text{del}$ (Gyr) & [2, 10] & 9.37$^{+0.25}_{-0.21}$ & 9.12$^{+1.84}_{-0.80}$ & 11.22$^{+1.08}_{-1.45}$\\
    log10(M$_\text{del}$/$M_{\odot}$) & [6, 10] & 8.11$^{+0.21}_{-0.25}$  & 8.52$^{+0.06}_{-0.07}$ & 8.52$^{+0.13}_{-0.15}$\\
    Age$_\text{exp}$ (Gyr) & [0.01, 4] & 0.34 $\pm$ 0.01 & 0.23 $\pm$ 0.02 & 0.54$^{+0.11}_{-0.09}$\\
    log10(M$_\text{exp}$/$M_{\odot}$) & [6, 10] & 8.40 $\pm$ 0.02 & 7.94 $\pm$ 0.04 & 7.98$^{+0.10}_{-0.17}$\\
    $\tau_{\text{exp}}$ (Gyr) & [0.01, 0.35] & 0.05 $\pm$ 0.004 & 0.03 $\pm$ 0.004 & 0.05$^{+0.02}_{-0.01}$\\
    $Z_{\text{exp}}$ & [0.1, 2.0] & 2.30$^{+0.12}_{-0.15}$ & 0.16$^{+0.07}_{-0.03}$ & 1.03$^{+0.38}_{-0.34}$\\
    logU & [-4.0, -2.0] & -3.45 $\pm$ 0.24 & -2.64 $^{+0.09}_{-0.08}$& -2.96$^{+0.60}_{-0.66}$\\
    $z$ & [0.19, 0.23] & 0.21 $\pm$ 0.005 & 0.21 $\pm$ 0.005 & 0.21 $\pm$ 0.005\\
    $\sigma_{\text{vel}}$ & [200, 600] & 401.04$^{+30.18}_{-32.26}$ & 435.05$^{+30.60}_{-30.24}$ & 531.72$^{+77.31}_{-67.36}$\\
    $A_{V_{\text{dust}}}$ & [0.0, 2.0] & 0.53 $\pm$ 0.03 & 0.31 $\pm$ 0.03 & 0.60 $\pm$ 0.16
    \enddata
\end{deluxetable*}
\subsection{Evidence of Extreme Ionization History in ASASSN-14li's Host}

\begin{figure}
    \centering
    \includegraphics[scale=0.31]{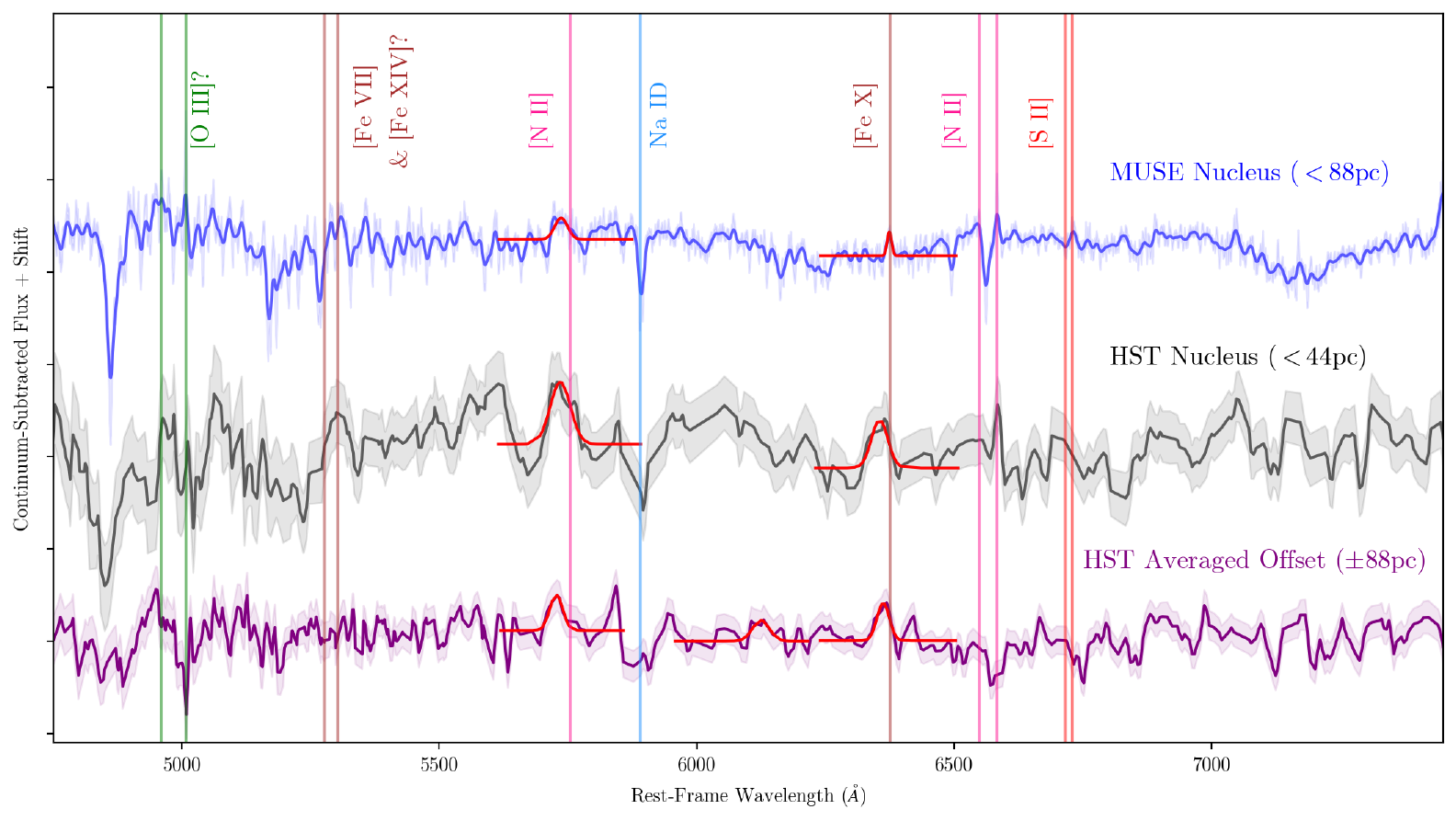}
    \caption{The nuclear spectrum of ASASSN-14li (with an extraction radius within 44 pc of the host center, in black) and the averaged spectrum from the extractions centered at $\pm$88 pc (purple), both smoothed to show the emission lines of ionized gas either detected in galaxy-integrated ground-based spectra or detected in this nuclear spectrum. We also include the MUSE spectrum of the same host at the smallest resolution radius of MUSE (0.2 arcsec) corresponding to light within 88pc of the SMBH, in blue. We label the locations of emission features (some not identified in any spectrum) with vertical lines, with $[$\ion{O}{3}$]$ $\lambda$$\lambda$4959,5007 in green, $[$\ion{Fe}{7}$]$ $\lambda$5276, $[$\ion{Fe}{14}$]$ $\lambda$5303, and $[$\ion{Fe}{10}$]$ $\lambda$6375 in brown, $[$\ion{N}{2}$]$ $\lambda$5755 and $[$\ion{N}{2}$]$ $\lambda\lambda$6548,6583 in dark pink, and $[$\ion{S}{2}$]$ $\lambda\lambda$ 6716, 6730 in red. Finally, we show the best-fit Gaussian to the $[$\ion{N}{2}$]$ $\lambda$5755 and $[$\ion{Fe}{10}$]$ $\lambda$6375 features, identified in all three spectra, in red overlays. These observations localize the high-ionization coronal lines to an emitting region closest to the host nucleus.}
    \label{fig:gaslines}
\end{figure}
The MUSE observations of ASASSN-14li's host \citep{Prieto2016} revealed extended nebular filaments and ionization regions around the host galaxy, suggesting an environment shaped by past AGN activity (or TDE activity) and possibly a recent merger event triggering either. While the MUSE observations also identified ionized emission lines from the nucleus, these lines were weak relative to the overall continuum, and the MUSE resolution of their nuclear spectrum includes light within a $\sim$400pc radius. Our observations probe an order of magnitude smaller than the MUSE results and show emission features at $[$\ion{N}{2}$]$ $\lambda$5755 and $[$\ion{Fe}{10}$]$ $\lambda$6375. There are also narrow features at $[$\ion{O}{3}$]$ $\lambda$$\lambda$4959,5007, but they are both immediately adjacent to a sharp absorption feature. Similarly, there are emission features at $[$\ion{Fe}{7}$]$ $\lambda$5276 and $[$\ion{Fe}{14}$]$ $\lambda$5303, but the errors at this region are very high due to the stitching between the two ends of G430L and G750L.

While many of these lines, and particularly $[$\ion{O}{3}$]$, have been detected in integrated spectra of this host galaxy and in the MUSE data of the nuclear region, we note that the red side of our spectrum corresponding to the G750L grating has residual fringing even after correcting the data with the flat, so we cannot rule out the possibility that some emission features that we recover may in fact be fringe features.

To estimate the robustness of each feature, we fit a Gaussian profile to the lines $[$\ion{N}{2}$]$ $\lambda$5755, $[$\ion{Fe}{7}$]$ $\lambda$6087, and $[$\ion{Fe}{10}$]$ $\lambda$6375 to estimate the total flux in each line above the local continuum, as shown in Figure \ref{fig:gaslines}. Using the per-pixel error on the flux, we measure the $[$\ion{Fe}{10}$]$ S/N to be $\sim$7, while the  S/N of both $[$\ion{N}{2}$]$ and $[$\ion{Fe}{7}$]$ is $\sim$16 in our HST nucleus observations. If these features are not fringing residuals, then the detections signify the localization of ionized gas to within 44 pc of the host nucleus. The averaged spectrum from the offset region at $\pm$88 pc only shows a comparable bump around $[$\ion{N}{2}$]$ $\lambda$5755, but it has S/N $\sim$ 2.

\subsection{Differences Between SDSS and HST Spectra for ASASSN-14ae, PTF09ge, and iPTF15af hosts}\label{sec:results_other3}

We present the nuclear spectra of the other three hosts observed in the same run as that of ASASSN-14li, corresponding to those of ASASSN-14ae, iPTF15af, and PTF09ge. These HST nuclear spectra are shown in Figure \ref{fig:asassn14li_nucleus_offsets} alongside the corresponding archival SDSS spectrum for each host. Due to the higher redshifts for each of these hosts, their spatial scale in the 52x0.2 STIS slit is too large physically to extract secondary regions distinct from a wide radius around the nucleus. Furthermore, the nuclear extractions for these three sources lack sufficient S/N to obtain confident results from \texttt{BAGPIPES} modeling, as done for ASASSN-14li's host.

For the host of ASASSN-14ae, we detect some absorption features in common with the archival SDSS spectrum, particularly H$\beta$, H$\gamma$, Mg I, and Ca H+K. It is therefore of note that we do \textit{not} see clearly the $[$\ion{O}{3}$]$ that is detected in the SDSS spectrum, nor the Na I D absorption feature at 5890 $\text{\AA}$ that is also clear in SDSS, although the latter could be due to the even lower S/N of the G750L observation and its fringing.

For the host of PTF09ge, we intriguingly detect Fe II $\lambda$5316 at S/N $\approx$ 3 and $[$\ion{Fe}{6}$]$ $\lambda$5677 (or \ion{N}{2} $\lambda$5679) at S/N $\approx$ 8 using Gaussian fitting. These lines are not detected in the SDSS archival spectra (S/N $\approx$ 1.5 for each). The presence of these lines at the nuclear region alongside the non-detection of the higher-ionization $[$\ion{Fe}{10}$]$ or $[$\ion{Fe}{7}$]$ lines -- and the lack of $[$\ion{O}{3}$]$ emission -- is the first known instance of such a configuration. The lines of $[$\ion{Fe}{6}$]$ at $\lambda$5146 (which we do not detect) and $\lambda$5677 are sensitive to the electron temperature of the emitting gas, but without other clear observed lines in the HST spectrum to discern the gas density, we cannot easily constrain the temperature. 

We do note that the $[$\ion{S}{2}$]$ $\lambda$$\lambda$6716, 6731 doublet is observed in both the SDSS and HST spectra for PTF09ge's host, but the lines are not resolved in the HST observation. We measure the $[$\ion{S}{2}$]$ $\lambda$6716/$[$\ion{S}{2}$]$$\lambda$6731 ratio from the SDSS spectrum to be $0.52^{0.58}_{0.41}$, indicative of gas with electrons denser than 10$^{3}$ cm$^{-3}$ \citep{OsterbrockFerland2006}, comparable to the electron densities of the innermost nuclear regions in nearby AGN \citep{2006A&A...456..953B} and among the highest sub-pc gas densities inferred from synchrotron afterglow modelling of TDE radio emission \citep[][]{Alexander2020}. Since this value is from a spectrum with light integrated within 1900 pc of the nucleus, we take this to be a likely lower limit on the density of the gas at the smaller scales probed by the HST observations. 

We then use this lower limit density (testing between $10^{3}$ up to $10^{9}$ cm$^{-3}$) and the measured strength of the $[$\ion{Fe}{6}$]$ $\lambda$5677 line (F$_{\text{[FeVI]}}$ = $3.19\times10^{-16}\pm 4.59\times 10^{-17}$ erg/s) in the Python package \texttt{PyNeb} \citep{PyNeb2015} across a range of temperatures (from $10^{2}$ to $10^{9}$ K) to estimate what temperature could explain the non-detection of other lines for PTF09ge's host. The upper-limit flux around the undetected line $[$\ion{Fe}{6}$]$ $\lambda$5146 (7.52 $\times 10^{-17}$ erg/s) can be used alongside the detected $[$\ion{Fe}{6}$]$ line, and we find the closest reproduced line ratios occur with temperatures closest to $10^{3}$ K at any density. However, no configuration reproduces the non-detection of any $[$\ion{Fe}{7}$]$ lines, which could be a result of the steady-state assumptions in \texttt{PyNeb}.
We do not favor the interpretation of the emission feature at $\lambda$5677 to be N II $\lambda$5677, as such N II lines have only been observed in contexts of O and Wolf-Rayet stars \citep[e.g.][]{1970MmRAS..73..153W} or in planetary nebulae \citep[e.g.][]{1976ApJS...31..517K}, and we identify no accompanying lines that would be expected from either context.

Finally, for iPTF15af, our highest-redshift host, we do not confidently detect any emission features, nor do we recover the Balmer or Mg absorption features seen in the SDSS spectrum of the same host. Increasing the extraction radius does not improve the S/N. We note that this target had shorter exposure times than the nearer host of PTF09ge, which likely further limited the depth of the HST spectra. As such, the non-detections in this case may reflect observational limitations rather than physical differences in the nuclear environment.

\section{Discussion} \label{sec:discussion}

Here we assess the significance of the different ages of the younger stellar populations as measured at the nucleus and the offset region at $\pm$88 pc for the host of ASASSN-14li, as well as the implications of the stellar masses found for our smallest discernible radius of the nucleus on the density profile expected for nuclear star clusters. We also address the possible detection of high-ionization emission lines in multiple hosts as a signature of either low-level AGN activity or the impacts of TDE energetics on circumnuclear environments.

\subsection{Stellar Age Gradients of ASASSN-14li Host}

The stellar population fits to the nucleus and offset regions of ASASSN-14li's host show similar mass profiles but different age estimates for the more recent starburst, consistent with stellar age gradients found among E+A galaxies in \cite{Pracy2013}. If the stellar population is indeed younger near the nucleus ($\sim$340 Myrs at 44 pc) than at greater distances \citep[$\sim$550 Myrs at 88 pc, consistent with the age inferred using the SDSS spectrum whose 3$\arcsec$-diameter aperture enclosed light from within 650pc;][]{French2017}, this gradient could reflect a history of inward gas migration due to a merger, during which older stars at larger radii were left behind as gas was funneled centrally by gravitational torques, subsequently sustaining later bursts of star formation closer to the nucleus \citep{2010ASPC..423..177B}.

Additionally, the enhanced TDE rate in PSBs is often linked to the steep central ($\sim$1-10 pc-scale) stellar density profiles observed in these systems \citep{Stone2016, Stone2018}, which may result from merger-triggered starbursts \citep[e.g.][]{2006ApJ...646L..33Y}, although \cite{Teboul2025} show that higher central densities may in fact lower TDE rates due to strong scatterings. A radial gradient in stellar ages could (if extended inside our resolution limit) explain the steep density profiles: as stars are formed in successive ``shells" via gas infall toward the center, younger, more massive stars are formed near the nucleus, increasing both the stellar mass density toward the center and possibly the chance of TDEs.  Likewise, if such age gradients extend to smaller scales, they may also support alternative explanations for TDE rate enhancements linked to differential relaxation of stars in different mass ranges \citep{Bortolas2022}.

\subsection{Density Profile of ASASSN-14li Host}

The stellar mass within 44pc of ASASSN-14li's nucleus as found by \texttt{Bagpipes} corresponds to a density of $\rho(44\text{pc})$ $\sim 4500 \pm 600$ M$_{\odot}$ / pc$^{3}$. By using the photometry for the central point-source component of the host as measured by \cite{French2020}, we can estimate the density down to a 30pc radius to probe even closer to the nuclear star cluster \citep[expected to dominate at $\sim$ 5pc;][]{Boker2004, Cote2006, Georgiev2014}. We use the synthetic photometry of the \texttt{Bagpipes} best-fit nuclear spectrum in the WFC3/F814W band for the mass-to-light (M/L) ratio and equate this value to the M/L ratio expected for the photometric point source in the same band. With this method we measure the stellar mass density to be $\rho(30\text{pc}) \sim 5900 \pm 800$ M$_{\odot}$ / pc$^{3}$.

The density of resolved NSCs at either their effective radii or within 5 pc ranges from $\sim$2,500--15,000 M$_{\odot}$ / pc$^{3}$ for $\sim$Milky Way-mass nucleated galaxies \citep[but can be much lower in galaxies with total stellar mass $<10^{9} M_{\odot}$;][]{Georgiev2016, Pechetti2020, Hannah2024, Hannah2025}. The density we measure at the photometrically resolved 30 pc radius is within this range, so the mass we attribute to a 30-pc region may in fact be largely centered with an even smaller region, potentially corresponding to an NSC. For the stellar mass of ASASSN-14li's host \citep[log(M$_{*}$) = 9.7;][]{French2016}, the results from \cite{Hannah2025} suggest that a NSC in this host would have a density range of $\rho(5\text{pc}) \sim 2600-4500$ M$_{\odot}$/pc$^{3}$. The surface brightness profile at small scales can extend steeply toward the center, especially in systems with compact NSCs. Thus, the high density at 30 pc may reflect the cumulative contribution of stellar populations at even smaller radii.

Indeed, \cite{French2020} found that this host was so centrally concentrated that, without a point-source component, the S\'ersic index was greater than 5 in F625W and F814W, and even exceeded 10 in F438W. The high density and strongly steeped brightness profile indicate we are likely inferring stellar population properties of the nuclear stellar cluster and differentiating its star formation history from the population at larger radii. This steep brightness profile is typical of PSBs; \cite{2008ApJ...688..945Y} found that most post-starburst galaxies (not yet known to have hosted a TDE) displayed similarly high S\'ersic indexes, indicating a commonality among PSBs for being uniquely centrally concentrated compared to elliptical counterparts. By evaluating the star formation histories within the central point-source region and just outside it in other PSBs (with and without known TDEs) we can confirm the association of inward star formation shells driven by gas inflows of a recent merger.

\subsection{Circumnuclear Ionized Gas in Hosts of ASASSN-14li, ASASSN-14ae, and PTF09ge}

The detection of $[$\ion{Fe}{10}$]$ $\lambda$6375 in the host of ASASSN-14li is significant because it is a high-ionization line that requires ionizing photons with energies exceeding 230 eV. Such lines are commonly associated with active galactic nuclei (AGN) \citep{Gelbord2009, 2011ApJ...739...69M} or extremely energetic transient processes like shocks from strong outflows or tidal disruption events (TDEs) \citep[e.g.][]{Komossa2008, Hinkle2023, Newsome2024_upjpaper}. Their presence within a 44 pc radius in ASASSN-14li's host suggests localized sources of intense ionizing radiation in the nuclear region, such as low-luminosity AGN activity.

We note that these lines could also have been ionized by the aftermath of the TDE itself, though the light crossing time between the source of the TDE emission and our smallest-resolution radius of 44 pc is $\sim$150 years. \cite{Makrygianni2025} found that there is a 132x enhancement factor in the TDE rate for galaxies with Lick H$\delta_{A}$ indices between 4.43 and 6, which includes ASASSN-14li's host \citep{French2016}. Using the global rate estimate of 3.2 $\times$ 10$^{-5}$ TDEs per galaxy per year from \cite{Yao2023}, we predict ASASSN-14li's host to have a TDE rate of 4.2 $\times$ 10$^{-3}$ events per year, or roughly one event every $\sim$250 years. Thus the ionization we see on the resolved scale may have been caused by a prior TDE. However, if the ionized lines indeed followed ASASSN-14li 2--3 years after its optical peak, then the gas would lie within the inner parsec of the nucleus. This could also explain the lack of $[$\ion{Fe}{7}$]$ $\lambda$6087 and $[$\ion{Fe}{10}$]$ at the offset region centered at $\pm$88 pc, but the signal is very limited at that distance.

The SDSS spectrum for ASASSN-14ae comes from light integrated within a $\sim$1250 pc region, whereas the HST nuclear spectrum of the same source is measured from light within a 66 pc radius of the SMBH. Thus the tentative non-detection of $[$\ion{O}{3}$]$ at the nucleus, similar to the findings for the host of ASASSN-14li, may be due to different energetics at different radial shells around the SMBH from a prior front of ionizing light passing through material over time. This could be a result of past TDE activity in the same host, which can produce AGN-like signatures at different locations throughout the host for $\sim10^{4}$ years based on the varying density of material at different radii from the central SMBH \citep{Mummery2025}. For instance, if the gas gets denser ($n_{H}\sim10^{9}$ cm$^{-3}$) as we approach closer to the nucleus due to gas inflows, then the emission from $[$\ion{Fe}{10}$]$ would dominate over any underlying co-existing $[$\ion{O}{3}$]$ emission. 

If these hosts have indeed experienced gas inflow and turbulence from a minor merger in the past, then the enhanced gas density at the center \textit{and} increased rate of TDE activity expected from the inflow could subsequently generate the suppression in $[$\ion{O}{3}$]$, especially when $[$\ion{Fe}{10}$]$ \textit{is} observed. A circumnuclear environment with high gas density due to merger inflows can also produce the combination of emission line features seen galaxy-wide effects such as extended emission line regions and line ratios changing on $\sim$decades timescales (or $<$ decade for denser material closer to the SMBH which would also produce extreme coronal emission lines).

The presence of Fe II $\lambda$6316 and $[$\ion{Fe}{6}$]$ $\lambda$5677 emission features in the host of PTF09ge, which are not as clear in the archival SDSS spectrum of the same host, are a glimpse into the layered ionization energies surrounding TDE-hosting SMBHs. These lines indicate an underlying ionizing continuum of lower energy than that required for even ``lower"-energy coronal lines such as $[$\ion{Fe}{7}$]$. Their confident detection only in the HST small-scale nuclear spectrum implies that the innermost region has distinct physical conditions from the rest of the host. The nuclear region may host a compact, chemically enriched component that could be masked in the integrated light of the SDSS spectrum. Such a difference could be driven by localized processes, such as residual star formation, weak AGN activity, or past tidal disruption events that have modified the gas conditions.

We note as well that the temperature estimates of the gas inferred from the line detections in PTF09ge's host are based on the gas density measured from the SDSS spectrum's $[$\ion{S}{2}$]$, which likely constitutes a lower limit on the density closer to the nucleus, and thus the host of PTF09ge's nuclear gas conditions could be even more extreme. The formation of $[$\ion{Fe}{6}$]$ requires ionizing photons with energies $\geq$99 eV, consistent with a relatively soft but still high-energy ionizing continuum, and the critical density of $[$\ion{Fe}{6}$]$ \citep[10$^{7.6}$ cm$^{-3}$;][]{1988AJ.....95...45A} allows for a wide range of densities above our lowest estimate. The absence of $[$\ion{Fe}{7}$]$ in PTF09ge suggests either a truncated or filtered ionizing continuum lacking photons $\geq$125 eV, or localized conditions (e.g., shielding or density structure) that suppress higher-ionization lines.

Finally, the non-detection of both coronal and lower-ionization lines in the host of iPTF15af in the HST spectra may also indicate a dearth of dense gas in the nuclear region, in contrast with the other hosts observed; however, due to the low S/N of our observations of this host, we do not claim that the non-detections are physical.

\section{Conclusions}\label{sec:conclusion}

Our analysis of the nuclear region of ASASSN-14li’s host galaxy provides new insights into the relationship between stellar population gradients, nuclear stellar cluster density, and circumnuclear ionized gas in post-starburst environments. The radial age gradient, with younger stars closer to the nucleus, supports a scenario of merger-driven gas inflow and subsequent central star formation, potentially amplifying the stellar density near the SMBH and contributing to enhanced TDE rates. The measured stellar mass density at 30 pc ($\sim$5900 $M_{\odot}/$pc$^{3}$) suggests significant central concentration, likely dominated by an unresolved nuclear stellar cluster, as indicated by the steep brightness profile observed photometrically.

The high-ionization lines [\ion{Fe}{6}]$\lambda$5677, [\ion{Fe}{7}] $\lambda$6087, and [\ion{Fe}{10}] $\lambda$6375 found in ASASSN-14li and PTF09ge likely originate from TDE-driven outflows or residual AGN activity based on the required ionization energies \citep[but we cannot rule out that prior SNe shocks may have also produced the ionizing energy; e.g.][]{2016ApJ...826..150D}. In the case of ASASSN-14li's host, the lines may even originate from gas in the inner parsec. The lack of [\ion{O}{3}] on the smaller nuclear scale of ASASSN-14ae's host, despite being present in the SDSS spectrum, is also indicative of varying ionization fronts at different radii from the nucleus, underscoring the importance of considering time-variable ionization in TDE host nuclei---emission lines detected years after a TDE may not trace ongoing steady-state AGN activity, but may instead represent echoes of past ionization fronts propagating through the circumnuclear medium.

Between the stellar population gradients most clearly seen in ASASSN-14li, coronal emission lines present in both ASASSN-14li and PTF09ge, and nuclear [\ion{O}{3}] suppression evident in ASASSN-14ae, we demonstrate across three hosts that TDE host galaxies share common signatures of merger-driven inflow and variable nuclear ionization, even if the details differ by system. These findings demonstrate how the unique environments of post-starburst galaxies create conditions conducive to enhanced TDE rates and reveal key connections between stellar dynamics, gas ionization, and SMBH interactions. Altogether we present strong test cases for analyzing the innermost regions of TDE hosts at multiple locations in order to discern star formation histories and signs of merger activity as sources for the enhanced TDE rate in post-starburst galaxies. \\\\

\vspace{1cm}

This work relied on observations made with the NASA/ESA Hubble Space Telescope and obtained via the Space Telescope Science Institute, which is operated by the Association of Universities for Research in Astronomy, Inc., under NASA contract NAS5-26555. These observations are associated with program HSTGO-14717.

M. N. thanks the LSST-DA Data Science Fellowship Program, which is funded by LSST-DA, the Brinson Foundation, the WoodNext Foundation, and the Research Corporation for Science Advancement Foundation; her participation in the program has benefited this work. M.N. also thanks Dr. Andrew Mummery for the fruitful discussions particularly regarding the effect of late-time TDE plateau on host studies.

I.A. acknowledges support from the European Research Council (ERC) under the European Union Horizon 2020 research and innovation program (grant No. 852097). 
K.D.F. acknowledges support from NSF grant AST–2206164. 
A.I.Z. acknowledges support in part from grant NASA ADAP \#80NSSC21K0988 as well as grant NSF PHY-2309135 to the Kavli Institute for Theoretical Physics (KITP).

N.C.S. gratefully acknowledges support from the Israel Science Foundation (Individual Research Grant 2414/23) and the Binational Science Foundation (grant Nos. 2019772 and 2020397).

\vspace{5mm}
\facilities{HST}
\software{astropy \citep{2013A&A...558A..33A,2018AJ....156..123A}, \texttt{BAGPIPES} \citep{Carnall2018}}

\bibliography{main}{}

\begin{thebibliography}{}
\expandafter\ifx\csname natexlab\endcsname\relax\def\natexlab#1{#1}\fi
\providecommand{\url}[1]{\href{#1}{#1}}
\providecommand{\dodoi}[1]{doi:~\href{http://doi.org/#1}{\nolinkurl{#1}}}
\providecommand{\doeprint}[1]{\href{http://ascl.net/#1}{\nolinkurl{http://ascl.net/#1}}}
\providecommand{\doarXiv}[1]{\href{https://arxiv.org/abs/#1}{\nolinkurl{https://arxiv.org/abs/#1}}}

\bibitem[{{Alexander} {et~al.}(2016){Alexander}, {Berger}, {Guillochon},
  {Zauderer}, \& {Williams}}]{Alexander2016}
{Alexander}, K.~D., {Berger}, E., {Guillochon}, J., {Zauderer}, B.~A., \&
  {Williams}, P.~K.~G. 2016, \apjl, 819, L25,
  \dodoi{10.3847/2041-8205/819/2/L25}

\bibitem[{{Alexander} {et~al.}(2020){Alexander}, {van Velzen}, {Horesh}, \&
  {Zauderer}}]{Alexander2020}
{Alexander}, K.~D., {van Velzen}, S., {Horesh}, A., \& {Zauderer}, B.~A. 2020,
  \ssr, 216, 81, \dodoi{10.1007/s11214-020-00702-w}

\bibitem[{{Appenzeller} \& {Oestreicher}(1988)}]{1988AJ.....95...45A}
{Appenzeller}, I., \& {Oestreicher}, R. 1988, \aj, 95, 45,
  \dodoi{10.1086/114611}

\bibitem[{{Arcavi} {et~al.}(2014){Arcavi}, {Gal-Yam}, {Sullivan}, {Pan},
  {Cenko}, {Horesh}, {Ofek}, {De Cia}, {Yan}, {Yang}, {Howell}, {Tal},
  {Kulkarni}, {Tendulkar}, {Tang}, {Xu}, {Sternberg}, {Cohen}, {Bloom},
  {Nugent}, {Kasliwal}, {Perley}, {Quimby}, {Miller}, {Theissen}, \&
  {Laher}}]{Arcavi2014}
{Arcavi}, I., {Gal-Yam}, A., {Sullivan}, M., {et~al.} 2014, \apj, 793, 38,
  \dodoi{10.1088/0004-637X/793/1/38}

\bibitem[{{Astropy Collaboration} {et~al.}(2013){Astropy Collaboration},
  {Robitaille}, {Tollerud}, {Greenfield}, {Droettboom}, {Bray}, {Aldcroft},
  {Davis}, {Ginsburg}, {Price-Whelan}, {Kerzendorf}, {Conley}, {Crighton},
  {Barbary}, {Muna}, {Ferguson}, {Grollier}, {Parikh}, {Nair}, {Unther},
  {Deil}, {Woillez}, {Conseil}, {Kramer}, {Turner}, {Singer}, {Fox}, {Weaver},
  {Zabalza}, {Edwards}, {Azalee Bostroem}, {Burke}, {Casey}, {Crawford},
  {Dencheva}, {Ely}, {Jenness}, {Labrie}, {Lim}, {Pierfederici}, {Pontzen},
  {Ptak}, {Refsdal}, {Servillat}, \& {Streicher}}]{2013A&A...558A..33A}
{Astropy Collaboration}, {Robitaille}, T.~P., {Tollerud}, E.~J., {et~al.} 2013,
  \aap, 558, A33, \dodoi{10.1051/0004-6361/201322068}

\bibitem[{{Astropy Collaboration} {et~al.}(2018){Astropy Collaboration},
  {Price-Whelan}, {Sip{\H{o}}cz}, {G{\"u}nther}, {Lim}, {Crawford}, {Conseil},
  {Shupe}, {Craig}, {Dencheva}, {Ginsburg}, {VanderPlas}, {Bradley},
  {P{\'e}rez-Su{\'a}rez}, {de Val-Borro}, {Aldcroft}, {Cruz}, {Robitaille},
  {Tollerud}, {Ardelean}, {Babej}, {Bach}, {Bachetti}, {Bakanov}, {Bamford},
  {Barentsen}, {Barmby}, {Baumbach}, {Berry}, {Biscani}, {Boquien}, {Bostroem},
  {Bouma}, {Brammer}, {Bray}, {Breytenbach}, {Buddelmeijer}, {Burke},
  {Calderone}, {Cano Rodr{\'\i}guez}, {Cara}, {Cardoso}, {Cheedella}, {Copin},
  {Corrales}, {Crichton}, {D'Avella}, {Deil}, {Depagne}, {Dietrich}, {Donath},
  {Droettboom}, {Earl}, {Erben}, {Fabbro}, {Ferreira}, {Finethy}, {Fox},
  {Garrison}, {Gibbons}, {Goldstein}, {Gommers}, {Greco}, {Greenfield},
  {Groener}, {Grollier}, {Hagen}, {Hirst}, {Homeier}, {Horton}, {Hosseinzadeh},
  {Hu}, {Hunkeler}, {Ivezi{\'c}}, {Jain}, {Jenness}, {Kanarek}, {Kendrew},
  {Kern}, {Kerzendorf}, {Khvalko}, {King}, {Kirkby}, {Kulkarni}, {Kumar},
  {Lee}, {Lenz}, {Littlefair}, {Ma}, {Macleod}, {Mastropietro}, {McCully},
  {Montagnac}, {Morris}, {Mueller}, {Mumford}, {Muna}, {Murphy}, {Nelson},
  {Nguyen}, {Ninan}, {N{\"o}the}, {Ogaz}, {Oh}, {Parejko}, {Parley}, {Pascual},
  {Patil}, {Patil}, {Plunkett}, {Prochaska}, {Rastogi}, {Reddy Janga},
  {Sabater}, {Sakurikar}, {Seifert}, {Sherbert}, {Sherwood-Taylor}, {Shih},
  {Sick}, {Silbiger}, {Singanamalla}, {Singer}, {Sladen}, {Sooley},
  {Sornarajah}, {Streicher}, {Teuben}, {Thomas}, {Tremblay}, {Turner},
  {Terr{\'o}n}, {van Kerkwijk}, {de la Vega}, {Watkins}, {Weaver}, {Whitmore},
  {Woillez}, {Zabalza}, \& {Astropy Contributors}}]{2018AJ....156..123A}
{Astropy Collaboration}, {Price-Whelan}, A.~M., {Sip{\H{o}}cz}, B.~M., {et~al.}
  2018, \aj, 156, 123, \dodoi{10.3847/1538-3881/aabc4f}

\bibitem[{{Bennert} {et~al.}(2006){Bennert}, {Jungwiert}, {Komossa}, {Haas}, \&
  {Chini}}]{2006A&A...456..953B}
{Bennert}, N., {Jungwiert}, B., {Komossa}, S., {Haas}, M., \& {Chini}, R. 2006,
  \aap, 456, 953, \dodoi{10.1051/0004-6361:20065319}

\bibitem[{{Blagorodnova} {et~al.}(2019){Blagorodnova}, {Cenko}, {Kulkarni},
  {Arcavi}, {Bloom}, {Duggan}, {Filippenko}, {Fremling}, {Horesh},
  {Hosseinzadeh}, {Karamehmetoglu}, {Levan}, {Masci}, {Nugent}, {Pasham},
  {Veilleux}, {Walters}, {Yan}, \& {Zheng}}]{Blagorodnova2019}
{Blagorodnova}, N., {Cenko}, S.~B., {Kulkarni}, S.~R., {et~al.} 2019, \apj,
  873, 92, \dodoi{10.3847/1538-4357/ab04b0}

\bibitem[{{B{\"o}ker} {et~al.}(2004){B{\"o}ker}, {Sarzi}, {McLaughlin}, {van
  der Marel}, {Rix}, {Ho}, \& {Shields}}]{Boker2004}
{B{\"o}ker}, T., {Sarzi}, M., {McLaughlin}, D.~E., {et~al.} 2004, \aj, 127,
  105, \dodoi{10.1086/380231}

\bibitem[{{Bortolas}(2022)}]{Bortolas2022}
{Bortolas}, E. 2022, \mnras, 511, 2885, \dodoi{10.1093/mnras/stac262}

\bibitem[{{Bournaud}(2010)}]{2010ASPC..423..177B}
{Bournaud}, F. 2010, in Astronomical Society of the Pacific Conference Series,
  Vol. 423, Galaxy Wars: Stellar Populations and Star Formation in Interacting
  Galaxies, ed. B.~{Smith}, J.~{Higdon}, S.~{Higdon}, \& N.~{Bastian}, 177,
  \dodoi{10.48550/arXiv.0909.1812}

\bibitem[{{Bruzual} \& {Charlot}(2003)}]{BC03}
{Bruzual}, G., \& {Charlot}, S. 2003, \mnras, 344, 1000,
  \dodoi{10.1046/j.1365-8711.2003.06897.x}

\bibitem[{{Calzetti} {et~al.}(2000){Calzetti}, {Armus}, {Bohlin}, {Kinney},
  {Koornneef}, \& {Storchi-Bergmann}}]{Calzetti2000}
{Calzetti}, D., {Armus}, L., {Bohlin}, R.~C., {et~al.} 2000, \apj, 533, 682,
  \dodoi{10.1086/308692}

\bibitem[{{Carnall} {et~al.}(2018){Carnall}, {McLure}, {Dunlop}, \&
  {Dav{\'e}}}]{Carnall2018}
{Carnall}, A.~C., {McLure}, R.~J., {Dunlop}, J.~S., \& {Dav{\'e}}, R. 2018,
  \mnras, 480, 4379, \dodoi{10.1093/mnras/sty2169}

\bibitem[{{Cendes} {et~al.}(2024){Cendes}, {Berger}, {Alexander}, {Chornock},
  {Margutti}, {Metzger}, {Wieringa}, {Bietenholz}, {Hajela}, {Laskar}, {Stroh},
  \& {Terreran}}]{Cendes2024}
{Cendes}, Y., {Berger}, E., {Alexander}, K.~D., {et~al.} 2024, \apj, 971, 185,
  \dodoi{10.3847/1538-4357/ad5541}

\bibitem[{{Chabrier}(2003)}]{Chabrier2003}
{Chabrier}, G. 2003, \pasp, 115, 763, \dodoi{10.1086/376392}

\bibitem[{{Clark} {et~al.}(2024){Clark}, {Graur}, {Callow}, {Aguilar}, {Ahlen},
  {Anderson}, {Berger}, {M{\"u}ller-Bravo}, {Brink}, {Brooks}, {Chen},
  {Claybaugh}, {de la Macorra}, {Doel}, {Filippenko}, {Forero-Romero}, {Gomez},
  {Gromadzki}, {Honscheid}, {Inserra}, {Kisner}, {Landriau}, {Makrygianni},
  {Manera}, {Meisner}, {Miquel}, {Moustakas}, {Nicholl}, {Nie}, {Onori},
  {Palmese}, {Poppett}, {Reynolds}, {Rezaie}, {Rossi}, {Sanchez}, {Schubnell},
  {Tarl{\'e}}, {Weaver}, {Wevers}, {Young}, {Zheng}, \& {Zhou}}]{Clark2023}
{Clark}, P., {Graur}, O., {Callow}, J., {et~al.} 2024, \mnras, 528, 7076,
  \dodoi{10.1093/mnras/stae460}

\bibitem[{{C{\^o}t{\'e}} {et~al.}(2006){C{\^o}t{\'e}}, {Piatek}, {Ferrarese},
  {Jord{\'a}n}, {Merritt}, {Peng}, {Ha{\c{s}}egan}, {Blakeslee}, {Mei}, {West},
  {Milosavljevi{\'c}}, \& {Tonry}}]{Cote2006}
{C{\^o}t{\'e}}, P., {Piatek}, S., {Ferrarese}, L., {et~al.} 2006, \apjs, 165,
  57, \dodoi{10.1086/504042}

\bibitem[{{Dopita} {et~al.}(2016){Dopita}, {Seitenzahl}, {Sutherland}, {Vogt},
  {Winkler}, \& {Blair}}]{2016ApJ...826..150D}
{Dopita}, M.~A., {Seitenzahl}, I.~R., {Sutherland}, R.~S., {et~al.} 2016, \apj,
  826, 150, \dodoi{10.3847/0004-637X/826/2/150}

\bibitem[{{Dressler} \& {Gunn}(1983)}]{DresslerGunn1983}
{Dressler}, A., \& {Gunn}, J.~E. 1983, \apj, 270, 7, \dodoi{10.1086/161093}

\bibitem[{{French} {et~al.}(2016){French}, {Arcavi}, \&
  {Zabludoff}}]{French2016}
{French}, K.~D., {Arcavi}, I., \& {Zabludoff}, A. 2016, \apjl, 818, L21,
  \dodoi{10.3847/2041-8205/818/1/L21}

\bibitem[{{French} {et~al.}(2017){French}, {Arcavi}, \&
  {Zabludoff}}]{French2017}
---. 2017, \apj, 835, 176, \dodoi{10.3847/1538-4357/835/2/176}

\bibitem[{{French} {et~al.}(2020{\natexlab{a}}){French}, {Arcavi}, {Zabludoff},
  {Stone}, {Hiramatsu}, {van Velzen}, {McCully}, \& {Jiang}}]{French2020}
{French}, K.~D., {Arcavi}, I., {Zabludoff}, A.~I., {et~al.} 2020{\natexlab{a}},
  \apj, 891, 93, \dodoi{10.3847/1538-4357/ab7450}

\bibitem[{{French} {et~al.}(2020{\natexlab{b}}){French}, {Wevers}, {Law-Smith},
  {Graur}, \& {Zabludoff}}]{French2020b}
{French}, K.~D., {Wevers}, T., {Law-Smith}, J., {Graur}, O., \& {Zabludoff},
  A.~I. 2020{\natexlab{b}}, \ssr, 216, 32, \dodoi{10.1007/s11214-020-00657-y}

\bibitem[{{Gallazzi} {et~al.}(2005){Gallazzi}, {Charlot}, {Brinchmann},
  {White}, \& {Tremonti}}]{Gallazzi2005}
{Gallazzi}, A., {Charlot}, S., {Brinchmann}, J., {White}, S. D.~M., \&
  {Tremonti}, C.~A. 2005, \mnras, 362, 41,
  \dodoi{10.1111/j.1365-2966.2005.09321.x}

\bibitem[{{Gelbord} {et~al.}(2009){Gelbord}, {Mullaney}, \&
  {Ward}}]{Gelbord2009}
{Gelbord}, J.~M., {Mullaney}, J.~R., \& {Ward}, M.~J. 2009, \mnras, 397, 172,
  \dodoi{10.1111/j.1365-2966.2009.14961.x}

\bibitem[{{Georgiev} {et~al.}(2016){Georgiev}, {B{\"o}ker}, {Leigh},
  {L{\"u}tzgendorf}, \& {Neumayer}}]{Georgiev2016}
{Georgiev}, I.~Y., {B{\"o}ker}, T., {Leigh}, N., {L{\"u}tzgendorf}, N., \&
  {Neumayer}, N. 2016, \mnras, 457, 2122, \dodoi{10.1093/mnras/stw093}

\bibitem[{Georgiev \& Böker(2014)}]{Georgiev2014}
Georgiev, I.~Y., \& Böker, T. 2014, Monthly Notices of the Royal Astronomical
  Society, 441, 3570, \dodoi{10.1093/mnras/stu797}

\bibitem[{{Gezari}(2021)}]{Gezari2021}
{Gezari}, S. 2021, \araa, 59, 21, \dodoi{10.1146/annurev-astro-111720-030029}

\bibitem[{{Gezari} {et~al.}(2012){Gezari}, {Chornock}, {Rest}, {Huber},
  {Forster}, {Berger}, {Challis}, {Neill}, {Martin}, {Heckman}, {Lawrence},
  {Norman}, {Narayan}, {Foley}, {Marion}, {Scolnic}, {Chomiuk}, {Soderberg},
  {Smith}, {Kirshner}, {Riess}, {Smartt}, {Stubbs}, {Tonry}, {Wood-Vasey},
  {Burgett}, {Chambers}, {Grav}, {Heasley}, {Kaiser}, {Kudritzki}, {Magnier},
  {Morgan}, \& {Price}}]{Gezari2012}
{Gezari}, S., {Chornock}, R., {Rest}, A., {et~al.} 2012, \nat, 485, 217,
  \dodoi{10.1038/nature10990}

\bibitem[{{Graur} {et~al.}(2018){Graur}, {French}, {Zahid}, {Guillochon},
  {Mandel}, {Auchettl}, \& {Zabludoff}}]{Graur2018}
{Graur}, O., {French}, K.~D., {Zahid}, H.~J., {et~al.} 2018, \apj, 853, 39,
  \dodoi{10.3847/1538-4357/aaa3fd}

\bibitem[{{Guolo} \& {Mummery}(2025)}]{GuoloMummery2025}
{Guolo}, M., \& {Mummery}, A. 2025, \apj, 978, 167,
  \dodoi{10.3847/1538-4357/ad990a}

\bibitem[{{Hammerstein} {et~al.}(2021){Hammerstein}, {Gezari}, {van Velzen},
  {Cenko}, {Roth}, {Ward}, {Frederick}, {Hung}, {Graham}, {Foley}, {Bellm},
  {Cannella}, {Drake}, {Kupfer}, {Laher}, {Mahabal}, {Masci}, {Riddle},
  {Rojas-Bravo}, \& {Smith}}]{Hammerstein2021}
{Hammerstein}, E., {Gezari}, S., {van Velzen}, S., {et~al.} 2021, \apjl, 908,
  L20, \dodoi{10.3847/2041-8213/abdcb4}

\bibitem[{{Hammerstein} {et~al.}(2023){Hammerstein}, {van Velzen}, {Gezari},
  {Cenko}, {Yao}, {Ward}, {Frederick}, {Villanueva}, {Somalwar}, {Graham},
  {Kulkarni}, {Stern}, {Andreoni}, {Bellm}, {Dekany}, {Dhawan}, {Drake},
  {Fremling}, {Gatkine}, {Groom}, {Ho}, {Kasliwal}, {Karambelkar}, {Kool},
  {Masci}, {Medford}, {Perley}, {Purdum}, {van Roestel}, {Sharma}, {Sollerman},
  {Taggart}, \& {Yan}}]{Hammerstein2023}
{Hammerstein}, E., {van Velzen}, S., {Gezari}, S., {et~al.} 2023, \apj, 942, 9,
  \dodoi{10.3847/1538-4357/aca283}

\bibitem[{{Hannah} {et~al.}(2024){Hannah}, {Seth}, {Stone}, \& {van
  Velzen}}]{Hannah2024}
{Hannah}, C.~H., {Seth}, A.~C., {Stone}, N.~C., \& {van Velzen}, S. 2024, \aj,
  168, 137, \dodoi{10.3847/1538-3881/ad630a}

\bibitem[{{Hannah} {et~al.}(2025){Hannah}, {Stone}, {Seth}, \& {van
  Velzen}}]{Hannah2025}
{Hannah}, C.~H., {Stone}, N.~C., {Seth}, A.~C., \& {van Velzen}, S. 2025, \apj,
  988, 29, \dodoi{10.3847/1538-4357/addd1b}

\bibitem[{{Hills}(1975)}]{1975Natur.254..295H}
{Hills}, J.~G. 1975, \nat, 254, 295, \dodoi{10.1038/254295a0}

\bibitem[{{Hinkle} {et~al.}(2024){Hinkle}, {Shappee}, \&
  {Holoien}}]{Hinkle2023}
{Hinkle}, J.~T., {Shappee}, B.~J., \& {Holoien}, T. W.~S. 2024, \mnras, 528,
  4775, \dodoi{10.1093/mnras/stae022}

\bibitem[{{Holoien} {et~al.}(2014){Holoien}, {Prieto}, {Bersier}, {Kochanek},
  {Stanek}, {Shappee}, {Grupe}, {Basu}, {Beacom}, {Brimacombe}, {Brown},
  {Davis}, {Jencson}, {Pojmanski}, \& {Szczygie{\l}}}]{Holoien2014}
{Holoien}, T.~W.~S., {Prieto}, J.~L., {Bersier}, D., {et~al.} 2014, \mnras,
  445, 3263, \dodoi{10.1093/mnras/stu1922}

\bibitem[{{Holoien} {et~al.}(2016){Holoien}, {Kochanek}, {Prieto}, {Stanek},
  {Dong}, {Shappee}, {Grupe}, {Brown}, {Basu}, {Beacom}, {Bersier},
  {Brimacombe}, {Danilet}, {Falco}, {Guo}, {Jose}, {Herczeg}, {Long},
  {Pojmanski}, {Simonian}, {Szczygie{\l}}, {Thompson}, {Thorstensen}, {Wagner},
  \& {Wo{\'z}niak}}]{Holoien2016}
{Holoien}, T.~W.~S., {Kochanek}, C.~S., {Prieto}, J.~L., {et~al.} 2016, \mnras,
  455, 2918, \dodoi{10.1093/mnras/stv2486}

\bibitem[{{Jiang} {et~al.}(2016){Jiang}, {Dou}, {Wang}, {Yang}, {Lyu}, \&
  {Zhou}}]{JiangN2016}
{Jiang}, N., {Dou}, L., {Wang}, T., {et~al.} 2016, \apjl, 828, L14,
  \dodoi{10.3847/2041-8205/828/1/L14}

\bibitem[{{Kaler}(1976)}]{1976ApJS...31..517K}
{Kaler}, J.~B. 1976, \apjs, 31, 517, \dodoi{10.1086/190390}

\bibitem[{{Kara} {et~al.}(2018){Kara}, {Dai}, {Reynolds}, \&
  {Kallman}}]{Kara2018}
{Kara}, E., {Dai}, L., {Reynolds}, C.~S., \& {Kallman}, T. 2018, \mnras, 474,
  3593, \dodoi{10.1093/mnras/stx3004}

\bibitem[{{Komossa} {et~al.}(2008){Komossa}, {Zhou}, {Wang}, {Ajello}, {Ge},
  {Greiner}, {Lu}, {Salvato}, {Saxton}, {Shan}, {Xu}, \& {Yuan}}]{Komossa2008}
{Komossa}, S., {Zhou}, H., {Wang}, T., {et~al.} 2008, \apjl, 678, L13,
  \dodoi{10.1086/588281}

\bibitem[{{Law-Smith} {et~al.}(2017){Law-Smith}, {Ramirez-Ruiz}, {Ellison}, \&
  {Foley}}]{LawSmith2017}
{Law-Smith}, J., {Ramirez-Ruiz}, E., {Ellison}, S.~L., \& {Foley}, R.~J. 2017,
  \apj, 850, 22, \dodoi{10.3847/1538-4357/aa94c7}

\bibitem[{{Luridiana} {et~al.}(2015){Luridiana}, {Morisset}, \&
  {Shaw}}]{PyNeb2015}
{Luridiana}, V., {Morisset}, C., \& {Shaw}, R.~A. 2015, \aap, 573, A42,
  \dodoi{10.1051/0004-6361/201323152}

\bibitem[{{Makrygianni} {et~al.}(2025){Makrygianni}, {Arcavi}, {Newsome},
  {Bandopadhyay}, {Coughlin}, {Linial}, {Mockler}, {Quataert}, {Nixon},
  {Godson}, {Pursiainen}, {Leloudas}, {French}, {Zitrin}, {Faris}, {Lam},
  {Horesh}, {Sfaradi}, {Fausnaugh}, {Nakar}, {Ackley}, {Andrews},
  {Charalampopoulos}, {Davies}, {Dgany}, {Dyer}, {Farah}, {Fender}, {Green},
  {Howell}, {Killestein}, {Koivisto}, {Lyman}, {McCully}, {Mitchell}, {Padilla
  Gonzalez}, {Rhodes}, {Sahu}, {Terreran}, \& {Warwick}}]{Makrygianni2025}
{Makrygianni}, L., {Arcavi}, I., {Newsome}, M., {et~al.} 2025, \apjl, 987, L20,
  \dodoi{10.3847/2041-8213/ade155}

\bibitem[{{Masterson} {et~al.}(2024){Masterson}, {De}, {Panagiotou}, {Kara},
  {Arcavi}, {Eilers}, {Frostig}, {Gezari}, {Grotova}, {Liu}, {Malyali},
  {Meisner}, {Merloni}, {Newsome}, {Rau}, {Simcoe}, \& {van
  Velzen}}]{Masterson2024}
{Masterson}, M., {De}, K., {Panagiotou}, C., {et~al.} 2024, \apj, 961, 211,
  \dodoi{10.3847/1538-4357/ad18bb}

\bibitem[{{Miller} {et~al.}(2015){Miller}, {Kaastra}, {Miller}, {Reynolds},
  {Brown}, {Cenko}, {Drake}, {Gezari}, {Guillochon}, {Gultekin}, {Irwin},
  {Levan}, {Maitra}, {Maksym}, {Mushotzky}, {O'Brien}, {Paerels}, {de Plaa},
  {Ramirez-Ruiz}, {Strohmayer}, \& {Tanvir}}]{2015Natur.526..542M}
{Miller}, J.~M., {Kaastra}, J.~S., {Miller}, M.~C., {et~al.} 2015, \nat, 526,
  542, \dodoi{10.1038/nature15708}

\bibitem[{{M{\"u}ller-S{\'a}nchez} {et~al.}(2011){M{\"u}ller-S{\'a}nchez},
  {Prieto}, {Hicks}, {Vives-Arias}, {Davies}, {Malkan}, {Tacconi}, \&
  {Genzel}}]{2011ApJ...739...69M}
{M{\"u}ller-S{\'a}nchez}, F., {Prieto}, M.~A., {Hicks}, E.~K.~S., {et~al.}
  2011, \apj, 739, 69, \dodoi{10.1088/0004-637X/739/2/69}

\bibitem[{{Mummery} {et~al.}(2025){Mummery}, {Guolo}, {Matthews}, {Newsome},
  {Lintott}, \& {Keel}}]{Mummery2025}
{Mummery}, A., {Guolo}, M., {Matthews}, J., {et~al.} 2025, arXiv e-prints,
  arXiv:2503.14163.
\newblock \doarXiv{2503.14163}

\bibitem[{{Mummery} {et~al.}(2024){Mummery}, {van Velzen}, {Nathan}, {Ingram},
  {Hammerstein}, {Fraser-Taliente}, \& {Balbus}}]{2024MNRAS.527.2452M}
{Mummery}, A., {van Velzen}, S., {Nathan}, E., {et~al.} 2024, \mnras, 527,
  2452, \dodoi{10.1093/mnras/stad3001}

\bibitem[{{Newsome} {et~al.}(2024{\natexlab{a}}){Newsome}, {Arcavi}, {Howell},
  {McCully}, {Terreran}, {Hosseinzadeh}, {Bostroem}, {Dgany}, {Farah}, {Faris},
  {Padilla-Gonzalez}, {Pellegrino}, \& {Andrews}}]{Newsome2024b}
{Newsome}, M., {Arcavi}, I., {Howell}, D.~A., {et~al.} 2024{\natexlab{a}},
  arXiv e-prints, arXiv:2406.11972, \dodoi{10.48550/arXiv.2406.11972}

\bibitem[{{Newsome} {et~al.}(2024{\natexlab{b}}){Newsome}, {Arcavi}, {Howell},
  {McCully}, {Terreran}, {Hosseinzadeh}, {Bostroem}, {Dgany}, {Farah}, {Faris},
  {Padilla-Gonzalez}, {Pellegrino}, \& {Andrews}}]{Newsome2024_upjpaper}
---. 2024{\natexlab{b}}, arXiv e-prints, arXiv:2406.11972,
  \dodoi{10.48550/arXiv.2406.11972}

\bibitem[{{Nicholl} {et~al.}(2022){Nicholl}, {Lanning}, {Ramsden}, {Mockler},
  {Lawrence}, {Short}, \& {Ridley}}]{Nicholl2022}
{Nicholl}, M., {Lanning}, D., {Ramsden}, P., {et~al.} 2022, \mnras, 515, 5604,
  \dodoi{10.1093/mnras/stac2206}

\bibitem[{{Osterbrock} \& {Ferland}(2006)}]{OsterbrockFerland2006}
{Osterbrock}, D.~E., \& {Ferland}, G.~J. 2006, {Astrophysics of gaseous nebulae
  and active galactic nuclei} (University Science Books)

\bibitem[{{Pasham} \& {van Velzen}(2018)}]{2018ApJ...856....1P}
{Pasham}, D.~R., \& {van Velzen}, S. 2018, \apj, 856, 1,
  \dodoi{10.3847/1538-4357/aab361}

\bibitem[{{Pechetti} {et~al.}(2020){Pechetti}, {Seth}, {Neumayer}, {Georgiev},
  {Kacharov}, \& {den Brok}}]{Pechetti2020}
{Pechetti}, R., {Seth}, A., {Neumayer}, N., {et~al.} 2020, \apj, 900, 32,
  \dodoi{10.3847/1538-4357/abaaa7}

\bibitem[{{Planck Collaboration} {et~al.}(2020){Planck Collaboration},
  {Aghanim}, {Akrami}, {Ashdown}, {Aumont}, {Baccigalupi}, {Ballardini},
  {Banday}, {Barreiro}, {Bartolo}, {Basak}, {Battye}, {Benabed}, {Bernard},
  {Bersanelli}, {Bielewicz}, {Bock}, {Bond}, {Borrill}, {Bouchet}, {Boulanger},
  {Bucher}, {Burigana}, {Butler}, {Calabrese}, {Cardoso}, {Carron},
  {Challinor}, {Chiang}, {Chluba}, {Colombo}, {Combet}, {Contreras}, {Crill},
  {Cuttaia}, {de Bernardis}, {de Zotti}, {Delabrouille}, {Delouis}, {Di
  Valentino}, {Diego}, {Dor{\'e}}, {Douspis}, {Ducout}, {Dupac}, {Dusini},
  {Efstathiou}, {Elsner}, {En{\ss}lin}, {Eriksen}, {Fantaye}, {Farhang},
  {Fergusson}, {Fernandez-Cobos}, {Finelli}, {Forastieri}, {Frailis},
  {Fraisse}, {Franceschi}, {Frolov}, {Galeotta}, {Galli}, {Ganga},
  {G{\'e}nova-Santos}, {Gerbino}, {Ghosh}, {Gonz{\'a}lez-Nuevo}, {G{\'o}rski},
  {Gratton}, {Gruppuso}, {Gudmundsson}, {Hamann}, {Handley}, {Hansen},
  {Herranz}, {Hildebrandt}, {Hivon}, {Huang}, {Jaffe}, {Jones}, {Karakci},
  {Keih{\"a}nen}, {Keskitalo}, {Kiiveri}, {Kim}, {Kisner}, {Knox},
  {Krachmalnicoff}, {Kunz}, {Kurki-Suonio}, {Lagache}, {Lamarre}, {Lasenby},
  {Lattanzi}, {Lawrence}, {Le Jeune}, {Lemos}, {Lesgourgues}, {Levrier},
  {Lewis}, {Liguori}, {Lilje}, {Lilley}, {Lindholm}, {L{\'o}pez-Caniego},
  {Lubin}, {Ma}, {Mac{\'\i}as-P{\'e}rez}, {Maggio}, {Maino}, {Mandolesi},
  {Mangilli}, {Marcos-Caballero}, {Maris}, {Martin}, {Martinelli},
  {Mart{\'\i}nez-Gonz{\'a}lez}, {Matarrese}, {Mauri}, {McEwen}, {Meinhold},
  {Melchiorri}, {Mennella}, {Migliaccio}, {Millea}, {Mitra},
  {Miville-Desch{\^e}nes}, {Molinari}, {Montier}, {Morgante}, {Moss}, {Natoli},
  {N{\o}rgaard-Nielsen}, {Pagano}, {Paoletti}, {Partridge}, {Patanchon},
  {Peiris}, {Perrotta}, {Pettorino}, {Piacentini}, {Polastri}, {Polenta},
  {Puget}, {Rachen}, {Reinecke}, {Remazeilles}, {Renzi}, {Rocha}, {Rosset},
  {Roudier}, {Rubi{\~n}o-Mart{\'\i}n}, {Ruiz-Granados}, {Salvati}, {Sandri},
  {Savelainen}, {Scott}, {Shellard}, {Sirignano}, {Sirri}, {Spencer},
  {Sunyaev}, {Suur-Uski}, {Tauber}, {Tavagnacco}, {Tenti}, {Toffolatti},
  {Tomasi}, {Trombetti}, {Valenziano}, {Valiviita}, {Van Tent}, {Vibert},
  {Vielva}, {Villa}, {Vittorio}, {Wandelt}, {Wehus}, {White}, {White},
  {Zacchei}, \& {Zonca}}]{2020A&A...641A...6P}
{Planck Collaboration}, {Aghanim}, N., {Akrami}, Y., {et~al.} 2020, \aap, 641,
  A6, \dodoi{10.1051/0004-6361/201833910}

\bibitem[{{Pracy} {et~al.}(2013){Pracy}, {Croom}, {Sadler}, {Couch},
  {Kuntschner}, {Bekki}, {Owers}, {Zwaan}, {Turner}, \& {Bergmann}}]{Pracy2013}
{Pracy}, M.~B., {Croom}, S., {Sadler}, E., {et~al.} 2013, \mnras, 432, 3131,
  \dodoi{10.1093/mnras/stt666}

\bibitem[{{Prieto} {et~al.}(2016){Prieto}, {Kr{\"u}hler}, {Anderson},
  {Galbany}, {Kochanek}, {Aquino}, {Brown}, {Dong}, {F{\"o}rster}, {Holoien},
  {Kuncarayakti}, {Maureira}, {Rosales-Ortega}, {S{\'a}nchez}, {Shappee}, \&
  {Stanek}}]{Prieto2016}
{Prieto}, J.~L., {Kr{\"u}hler}, T., {Anderson}, J.~P., {et~al.} 2016, \apjl,
  830, L32, \dodoi{10.3847/2041-8205/830/2/L32}

\bibitem[{{Rees}(1988)}]{Rees1988}
{Rees}, M.~J. 1988, \nat, 333, 523, \dodoi{10.1038/333523a0}

\bibitem[{{Sazonov} {et~al.}(2021){Sazonov}, {Gilfanov}, {Medvedev}, {Yao},
  {Khorunzhev}, {Semena}, {Sunyaev}, {Burenin}, {Lyapin}, {Meshcheryakov},
  {Uskov}, {Zaznobin}, {Postnov}, {Dodin}, {Belinski}, {Cherepashchuk},
  {Eselevich}, {Dodonov}, {Grokhovskaya}, {Kotov}, {Bikmaev}, {Zhuchkov},
  {Gumerov}, {van Velzen}, \& {Kulkarni}}]{Sazonov2021}
{Sazonov}, S., {Gilfanov}, M., {Medvedev}, P., {et~al.} 2021, \mnras, 508,
  3820, \dodoi{10.1093/mnras/stab2843}

\bibitem[{{Stone} {et~al.}(2018){Stone}, {Generozov}, {Vasiliev}, \&
  {Metzger}}]{Stone2018}
{Stone}, N.~C., {Generozov}, A., {Vasiliev}, E., \& {Metzger}, B.~D. 2018,
  \mnras, 480, 5060, \dodoi{10.1093/mnras/sty2045}

\bibitem[{{Stone} \& {van Velzen}(2016)}]{Stone2016}
{Stone}, N.~C., \& {van Velzen}, S. 2016, \apjl, 825, L14,
  \dodoi{10.3847/2041-8205/825/1/L14}

\bibitem[{{Teboul} \& {Perets}(2025)}]{Teboul2025}
{Teboul}, O., \& {Perets}, H.~B. 2025, \apj, 984, 12,
  \dodoi{10.3847/1538-4357/adc09f}

\bibitem[{{van Velzen} {et~al.}(2021){van Velzen}, {Pasham}, {Komossa}, {Yan},
  \& {Kara}}]{VanVelzen21}
{van Velzen}, S., {Pasham}, D.~R., {Komossa}, S., {Yan}, L., \& {Kara}, E.~A.
  2021, \ssr, 217, 63, \dodoi{10.1007/s11214-021-00835-6}

\bibitem[{{van Velzen} {et~al.}(2016){van Velzen}, {Anderson}, {Stone},
  {Fraser}, {Wevers}, {Metzger}, {Jonker}, {van der Horst}, {Staley}, {Mendez},
  {Miller-Jones}, {Hodgkin}, {Campbell}, \& {Fender}}]{2016Sci...351...62V}
{van Velzen}, S., {Anderson}, G.~E., {Stone}, N.~C., {et~al.} 2016, Science,
  351, 62, \dodoi{10.1126/science.aad1182}

\bibitem[{{Wackerling}(1970)}]{1970MmRAS..73..153W}
{Wackerling}, L.~R. 1970, \memras, 73, 153

\bibitem[{{Yang} {et~al.}(2006){Yang}, {Tremonti}, {Zabludoff}, \&
  {Zaritsky}}]{2006ApJ...646L..33Y}
{Yang}, Y., {Tremonti}, C.~A., {Zabludoff}, A.~I., \& {Zaritsky}, D. 2006,
  \apjl, 646, L33, \dodoi{10.1086/506909}

\bibitem[{{Yang} {et~al.}(2008){Yang}, {Zabludoff}, {Zaritsky}, \&
  {Mihos}}]{2008ApJ...688..945Y}
{Yang}, Y., {Zabludoff}, A.~I., {Zaritsky}, D., \& {Mihos}, J.~C. 2008, \apj,
  688, 945, \dodoi{10.1086/591656}

\bibitem[{{Yao} {et~al.}(2023){Yao}, {Ravi}, {Gezari}, {van Velzen}, {Lu},
  {Schulze}, {Somalwar}, {Kulkarni}, {Hammerstein}, {Nicholl}, {Graham},
  {Perley}, {Cenko}, {Stein}, {Ricarte}, {Chadayammuri}, {Quataert}, {Bellm},
  {Bloom}, {Dekany}, {Drake}, {Groom}, {Mahabal}, {Prince}, {Riddle},
  {Rusholme}, {Sharma}, {Sollerman}, \& {Yan}}]{Yao2023}
{Yao}, Y., {Ravi}, V., {Gezari}, S., {et~al.} 2023, \apjl, 955, L6,
  \dodoi{10.3847/2041-8213/acf216}

\end{thebibliography}
\bibliographystyle{aasjournal}

\appendix
\section{Star Formation History Fits}\label{cornerplots}

\begin{figure*}[h!]
    \centering
    \includegraphics[scale=0.25]{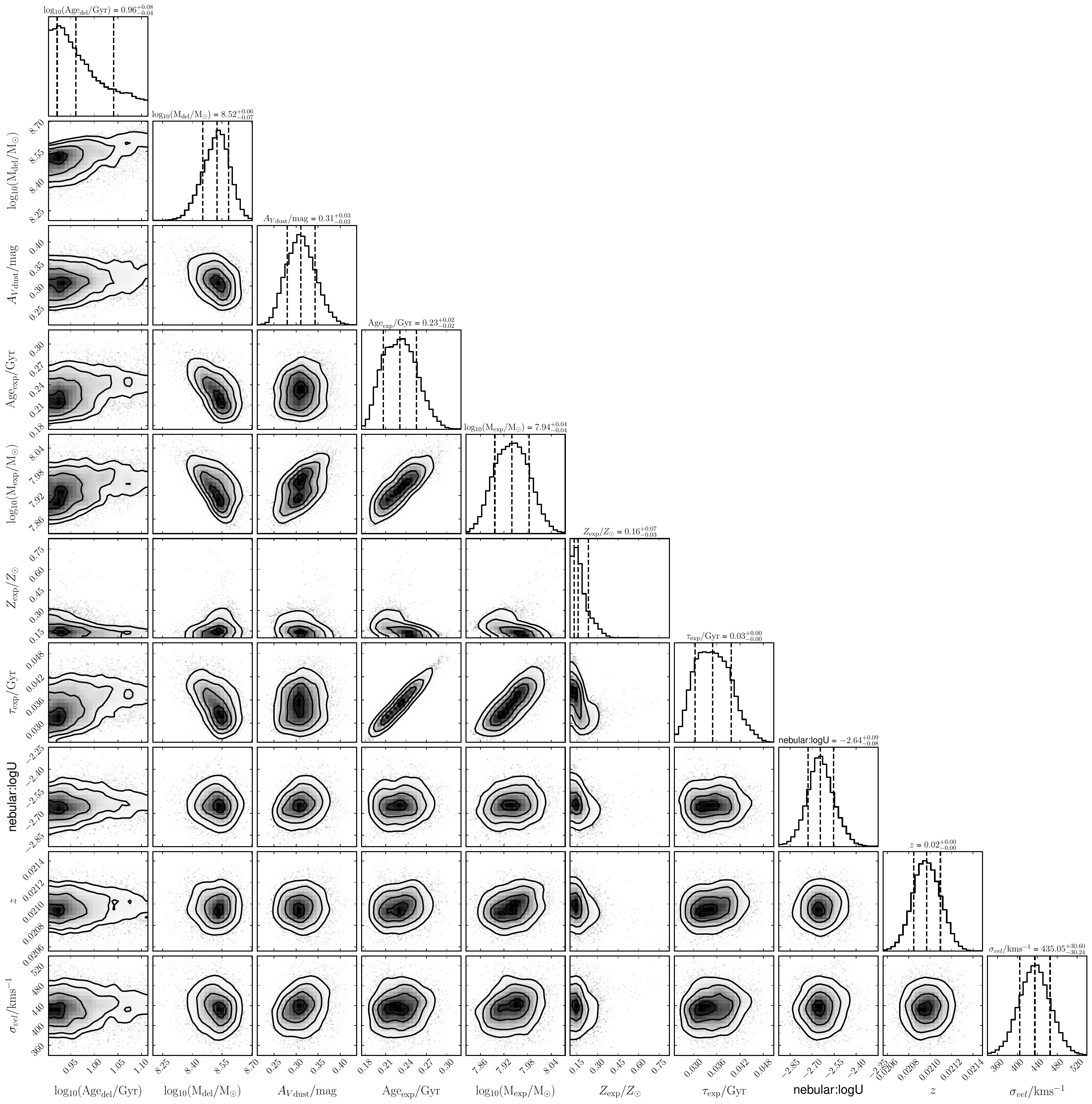}
    \caption{Corner plot showing 2D projection of the fitted parameter space for the nuclear spectrum (light within a 44pc radius from the nucleus) of ASASSN-14li's host galaxy. The calibration values (N0, N1, and N2) corresponding to the Chebyshev polynomial coefficients used to calibrate the flux have been omitted from the plot for visual clarity of the fits to physical parameters, but their best fits were centered at 1.06$\pm$0.15, 0.17$\pm$0.04, and 0.02$\pm$0.01, respectively.}
    \label{fig:bagpipes_corner}
\end{figure*}

\begin{figure*}
    \centering
    \includegraphics[scale=0.25]{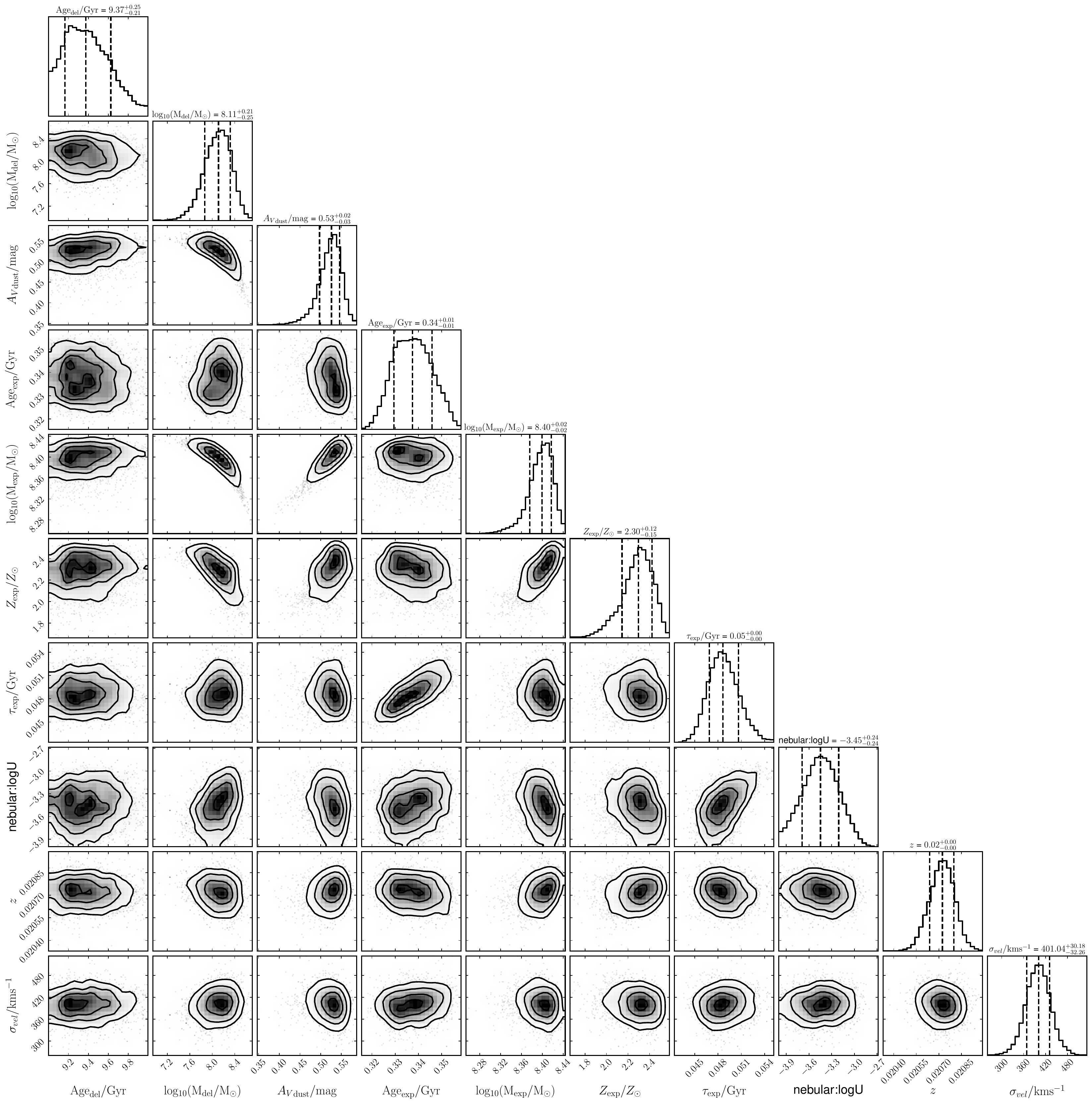}
    \caption{Same as Figure \ref{fig:bagpipes_corner_subtracted}, but with the estimated contribution from the central accretion disk (present at late times in ASASSN-14li, including at the time of our observations) removed from the nuclear spectrum. As in the prior figure, the calibration values (N0, N1, and N2) corresponding to the Chebyshev polynomial coefficients used to calibrate the flux have been omitted from the plot for visual clarity of the fits to physical parameters, but their best fits were centered at 1.10$\pm$0.16, -0.17$\pm$0.04, and 0.09$\pm$0.02, respectively. The values found from this fit to the disk-subtracted spectrum are used as the results for the nuclear region throughout this work.}
    \label{fig:bagpipes_corner_subtracted}
\end{figure*}

\begin{figure*}
    \centering
    \includegraphics[scale=0.31]{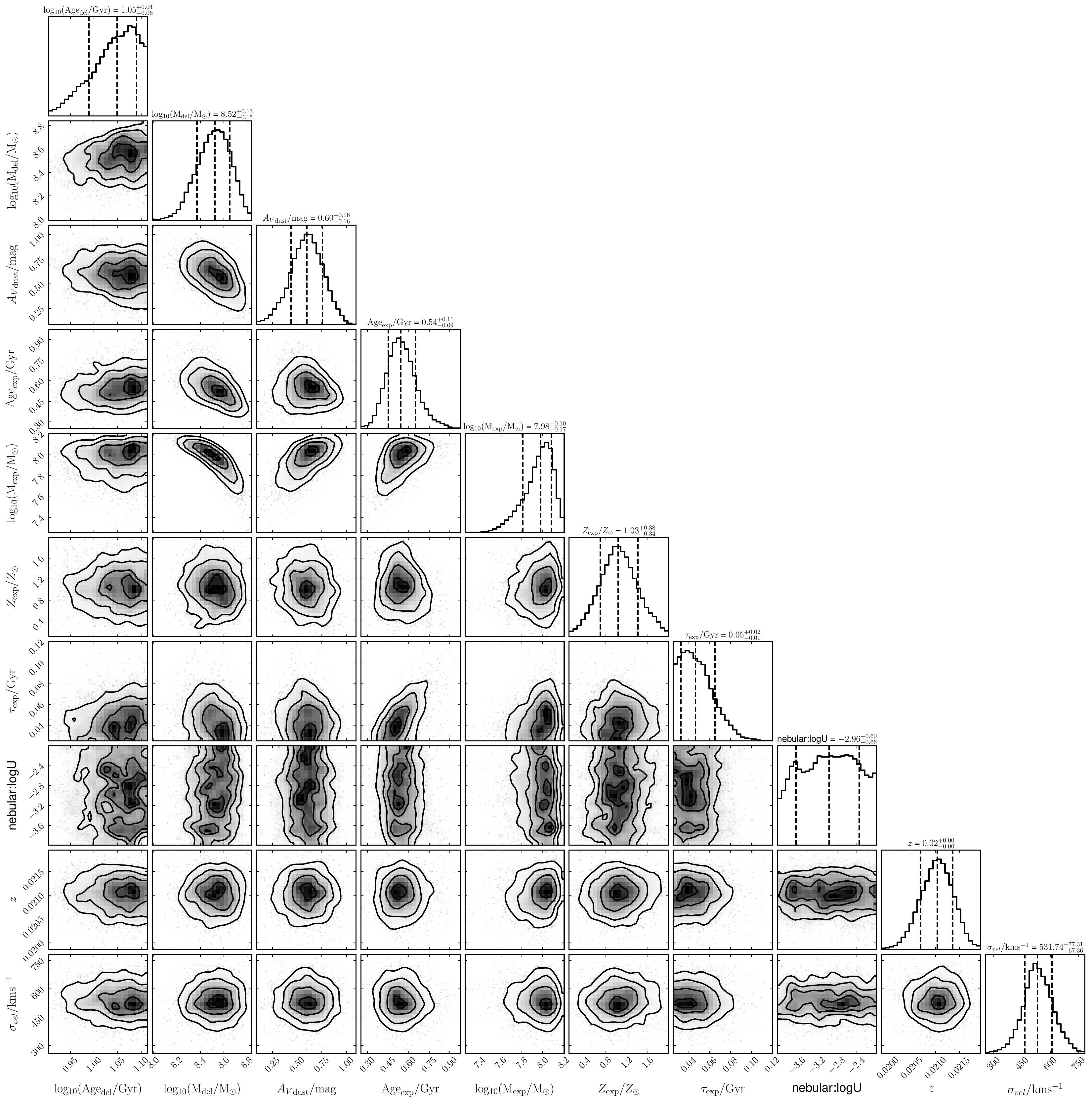}
    \caption{Same as Figures \ref{fig:bagpipes_corner} and \ref{fig:bagpipes_corner_subtracted}, but for the offset spectrum (light centered at an 88pc radius from the nucleus) of ASASSN-14li's host galaxy.}
    \label{fig:bagpipes_corner_offset}
\end{figure*}

Here, we provide the corner plot showing the convergence of fitted parameters for ASASSN-14li's innermost nuclear region using \texttt{BAGPIPES} in Figure \ref{fig:bagpipes_corner} (for the unsubtracted nuclear spectrum) and Figure \ref{fig:bagpipes_corner_subtracted} (for the nuclear spectrum with the accretion disk contribution from the TDE subtracted), as well as the corner plot of the \texttt{BAGPIPES} fit to the offset region in ASASSN-14li centered at $\pm$ 88 pc from the nucleus is shown in Figure \ref{fig:bagpipes_corner_offset}. The parameter ranges allowed for fitting and the derived properties found for the nuclear and offset regions are highlighted in Table \ref{table:bagpipes}.

\end{document}